\documentclass[useAMS,usenatbib]{mn2e}
\usepackage{graphicx}
 
\newcommand{\oversim}[2]{\protect{\mbox{\lower0.5ex\vbox{%
  \baselineskip=0pt\lineskip=0.2ex
  \ialign{$\mathsurround=0pt #1\hfil##\hfil$\crcr#2\crcr\sim\crcr}}}}}
\newcommand{\simgreat}{\mbox{$\,\mathrel{\mathpalette\oversim>}\,$}} 
\newcommand{\simless} {\mbox{$\,\mathrel{\mathpalette\oversim<}\,$}} 

\title[Maximum-stellar-mass--cluster-mass-relation]{The maximum
  stellar mass, star-cluster formation and composite stellar populations}

\author[C.~Weidner and P.~Kroupa]{Carsten Weidner$^{1,2}$ and
  Pavel Kroupa$^{1,2}$\thanks{e-mail:
    cweidner/pavel@astro.uni-bonn.de}\\ 
$^{1}$ Argelander Institut f\"ur Astronomie (Sternwarte), Universit\"at
Bonn, D-53121 Bonn, Germany\\
$^{2}$ The Rhine-Stellar-Dynamical Network}
\begin{document}

\date{Accepted 2005 November 3. Received 2005 October 29; in original
  form 2005 July 15}

\pagerange{\pageref{firstpage}--\pageref{lastpage}} \pubyear{2005}

\maketitle

\label{firstpage}
\begin{abstract}
We demonstrate that the mass of the most massive star in a cluster
correlates non-trivially with the cluster mass. A simple algorithm
according to which a cluster is filled up with stars that are chosen
randomly from the standard IMF but sorted with increasing mass yields
an excellent description of the observational data. Algorithms based
on random sampling from the IMF without sorted adding are ruled out
with a confidence larger than 0.9999. A physical explanation
of this would be that a cluster forms by more-massive stars being
consecutively added until the resulting feedback energy suffices to
revert cloud contraction and stops further star formation. This has
important implications for composite populations. For example,
$10^{4}$ clusters of mass $10^{2}\,M_{\odot}$ will not produce the
same IMF as one cluster with a mass of $10^{6}\,M_{\odot}$. It also
supports the notion that the integrated galaxial IMF (IGIMF) should be
steeper than the stellar IMF and that it should vary with the
star-formation rate of a galaxy.
\end{abstract}

\section{Introduction}
\label{sec:intro}

The insight that clustered star formation may be the dominant mode for
star formation has grown over the last years. The form of the true
distribution of stellar masses within these clusters, of the stellar
initial mass function (IMF), has been a subject of debate for a long
time. The 
evolution of the stars, unresolved binaries, and the dynamical
evolution of the clusters complicates the observational efforts to
extract the IMF. Unfortunately the most promising objects, very
young stellar clusters (age $<$ 3 Myr), are often still embedded in
their natal cloud - again aggravating observations. 

Nevertheless the distribution of stars in young clusters seems to be fairly
well described by a multi power-law function with a slope or index
($\alpha$) of 2.35 (the so-called 'Salpeter' value) for stars with a
mass larger than $0.5\,M_{\odot}$ \citep{Kr01}. The initial mass
function (IMF),
\begin{equation}
\xi(m) \propto m^{-\alpha_{i}}, 
\end{equation}
where $\xi(m)\,dm$ is the number of stars in the mass interval
$m,\,m+dm$.  
Several observations find the Salpeter value ($\alpha_{3} = 2.35$) for
a large variety of conditions
\citep{MH98,SND00,SND02,PaZa01,Mass02,Mass03,WGH02,BMK03,PBK04}. It is
therefore useful to describe the {\it stellar IMF} with an
invariant, multi-power law form \citep{KTG93,Kr01,RGH02},
\begin{equation}
          \begin{array}{l@{,\quad}l}
\alpha_0 = +0.30&0.01 \le m/{M}_\odot < 0.08,\\
\alpha_1 = +1.30&0.08 \le m/{M}_\odot < 0.50,\\
\alpha_2 = +2.35&0.50 \le m/{M}_\odot < 1.00,\\
\alpha_3 = +2.35&1.00 \le m/{M}_\odot.\\
          \end{array}
\label{Kroupa-IMF}
\end{equation}

\noindent
We refer to this form as the standard or canonical stellar IMF because
this form fits the luminosity function of Galactic-field and cluster
stars below 1 $M_{\odot}$ and also represents young populations above
1 $M_{\odot}$ \citep{KTG93,Kr01,Kr02}. {As pointed out by
\citet{Sc98,Sc04} though, significant uncertainties remain in the
determination of the IMF to the point that the case can also be made
that a single form of the IMF may not exist. In view of this, the
ansatz made here and elsewhere is to propose the hypothesis of an
invariant standard or canonical IMF (eq.~\ref{Kroupa-IMF}) and to test
if the variation of the observed IMF can be understood to be the
result of astrophysical effects (obscuration, stellar evolution,
stellar multiplicity), dynamical effects (mass segregation, stellar
evaporation and ejections), stochastic effects (finite $N$-sampling
from the IMF) and the construction of composite populations (addition
of many different clusters).}

Similarly, the embedded cluster mass function (ECMF) has been found to
be well-described by at least one power-law,
\begin{equation}
\xi_{\rm ecl}(M_{\rm ecl}) \propto M_{\rm ecl}^{-\beta}, 
\label{eq:ECMF}
\end{equation}
where $dN_{\rm ecl} = \xi_{\rm
  ecl}(M_{\rm ecl})~dM_{\rm ecl}$ is the number of embedded clusters
in the mass interval $M_{\rm ecl}$, $M_{\rm ecl} + dM_{\rm ecl}$ and
$M_{\rm ecl}$ is the cluster mass in stars. The observational evidence
points to a possibly universal form of the ECMF: \citet{LL03} find a
slope $\beta = 2$ in the solar neighbourhood for clusters with masses
between 50 and 1000 $M_{\odot}$, while \citet{HEDM03} find $2 \simless
\beta \simless 2.4$ for $10^{3} \simless M_{\rm ecl}/M_{\odot}
\simless 10^{4}$ in the SMC and LMC, and \citet{ZF99} find $1.95 \pm
0.03$ for $10^4 \simless M_{\rm ecl}/M_\odot \simless 10^6$ in the
Antennae galaxies. \citet{WKL04} discovered that $\beta = 2.35$ best
reproduces the observed correlation between the brightest young
cluster and the galaxy-wide star-formation-rate for a large sample of
late-type galaxies.

As already mentioned by \citet{VB82} and discussed in more detail by
\citet{KW03}, the composite or integrated galaxial stellar IMF (IGIMF) is
obtained by summing up the stellar IMFs contributed by all the star
clusters that formed over the age of a galaxy,
\begin{eqnarray}
\label{eq:IGIMF}
\xi_{\rm IGIMF}(m; t) &=& \int_{M_{\rm ecl,min}}^{M_{\rm ecl,max}(SFR(t))} \xi(m
\le
                      m_{\rm max}(M_{\rm ecl})) \nonumber \\
& &\cdot\,\xi_{\rm ecl}(M_{\rm ecl})\,dM_{\rm ecl},
\end{eqnarray}
where $\xi_{\rm ecl}(M_{\rm ecl})$ is the ECMF and $\xi(m\le m_{\rm
  max}(M_{\rm ecl}))$ is the stellar IMF in a particular cluster
within which the maximal mass of a star is $m_{\rm max}$. $M_{\rm
  ecl,min}$ (= $5\,M_{\odot}$, Taurus-Auriga-type ``clusters'') is the
minimal cluster mass, while the maximal cluster mass, $M_{\rm
  ecl,max}$, depends on the galaxy-wide star-formation rate, SFR
\citep{WKL04}. 

A critical function entering this description is thus
$m_{\rm max}(M_{\rm ecl})$. Assuming the stellar IMF to be a
continuous distribution function, this mass of the most massive star
in an embedded cluster with the total mass $M_{\rm ecl}$ in stars is
given by 
\begin{equation}
1 = \int_{m_{\rm max}}^{m_{\rm max*}} \xi(m)\,dm,
\label{eq:mmmc}
\end{equation}
with
\begin{equation}
M_{\rm ecl} = \int_{m_{\rm low}}^{m_{\rm max}} m\,\xi(m)\,dm,
\label{eq:Meclmc}
\end{equation}
since there exists exactly one most massive star in each cluster, and
neglecting statistical variations. Here $m_{\rm low}$ = 0.01 $M_{\odot}$
is the minimal fragmentation mass and $m_{\rm
  max*}\,\approx\,150\,M_{\odot}$ is the measured maximal stellar mass
limit \citep{WK04,Fi05,OC05}. On combining eqs.~\ref{eq:mmmc}
and~\ref{eq:Meclmc} the function
\begin{equation}
m_{\rm max} = {}^{l}m_{\rm max}^{\rm ana}(M_{\rm ecl})
\label{eq:MmaxvsMecl}
\end{equation}
is quantified by \citet{WK04} and in \S~\ref{sec:maxstmass}.  This is
the analytical (``ana'') maximum-stellar-mass--cluster-mass relation
which incorporates the fundamental stellar upper mass limit {(noted by
the leading superscript ``l'') of $m_{\rm max*}\,=\,150\,M_{\odot}$. 
Later-on other maximum-stellar-mass--cluster-mass relations are indicated by
different superscripts: ``ran'', ``con'' and ``sort'' for the
different Monte-Carlo sampling methods (see \ref{se:MC}) and also
``u'' for the case without a fundamental stellar upper mass limit.}

\citet{WK04} infer that a fundamental upper 
stellar mass limit, $m_{\rm max*}\,\approx\,150\,M_\odot$, appears to
exist above which stars do not occur, unless $\alpha_{3}\simgreat2.8$,
in which case no conclusions can be drawn based on the expected number
of massive stars. As reviewed by \citet{KW05}, the existence of such a
stellar upper mass limit has been further {substantiated} by
\citet{Fi05} and \citet{OC05} for a range of star clusters and OB
associations.

We thus have, for each $M_{\rm ecl}$, the maximal stellar mass,
$m_{\rm max}(M_{\rm ecl})\le m_{\rm max*}$, and with this information
eq.~\ref{eq:IGIMF} can be evaluated to compute the IGIMF. \citet{KW03}
find the IGIMF, when evaluated to the highest cluster masses, to be
significantly steeper than the stellar IMF, and \citet{WK05} extend
the analysis to a time varying ECMF by noting that $M_{\rm ecl, max}$
increases with the star-formation rate of a galaxy. They show the 
IGIMF to be not only steeper than the stellar IMF, but also to depend
on galaxy type. 
The implications of these findings are rather significant for the
supernova rate \citep{WK05} and for the chemical evolution of galaxies
\citep{KWK05}.

But these results remain not without a challenge. \citet{El04} argues
that there is no evidence of a relation $m_{\rm max} = m_{\rm max}(M_{\rm
  ecl}) \le m_{\rm max *}$. This relation implies that many small,
low-mass, star-forming events will not have the same combined IMF as
one major SF event of the same mass. Thus, according to \citet{KW05},
$10^{5}$ clusters each with a mass of $20\,M_{\odot}$ would provide a
combined IMF that differs from that of one cluster with a mass of
$2 \times 10^{6}\,M_{\odot}$ by being underrepresented in stars with a
mass above about $m_{\rm max}\,=\,1\,M_{\odot}$. The contrary, often
voiced view is that stellar masses sample the IMF purely statistically
such that $10^{5}$ clusters containing 50 stars (on average 20
$M_{\odot}$) will give the same combined IMF as one cluster containing $5
\times 10^{6}$ stars \citep{El99,El04}.

With this contribution we demonstrate conclusively that the purely
statistical notion is false, and that the stellar IMF is sampled to a
maximum stellar mass that correlates with the cluster mass. Therewith
we affirm the results obtained by \citet{KW03} and \citet{WK05}, and
we also attain useful insights into the process of star-cluster
formation.

In Section~\ref{sec:maxstmass} our Monte-Carlo procedure is described
and the maximal-star-mass--cluster-mass relation is derived, while in
\S~\ref{sec:montec} the Monte-Carlo experiment is applied to the IGIMF
and the results are presented. A discussion with conclusions is
available in \S~\ref{sec:concs}.
\section{The Maximal star mass in a cluster}
\label{sec:maxstmass}
\subsection{Previous' studies}
\label{se:PS}
Over the past 20 years several studies investigated a
possible connection of the maximum stellar mass in a cluster and the
mass of the cluster because such a relation, if it were to exist,
would allow important insights into the star-formation process.

\begin{figure}
\begin{center}
\vspace*{-0.6cm}
\includegraphics[width=8cm]{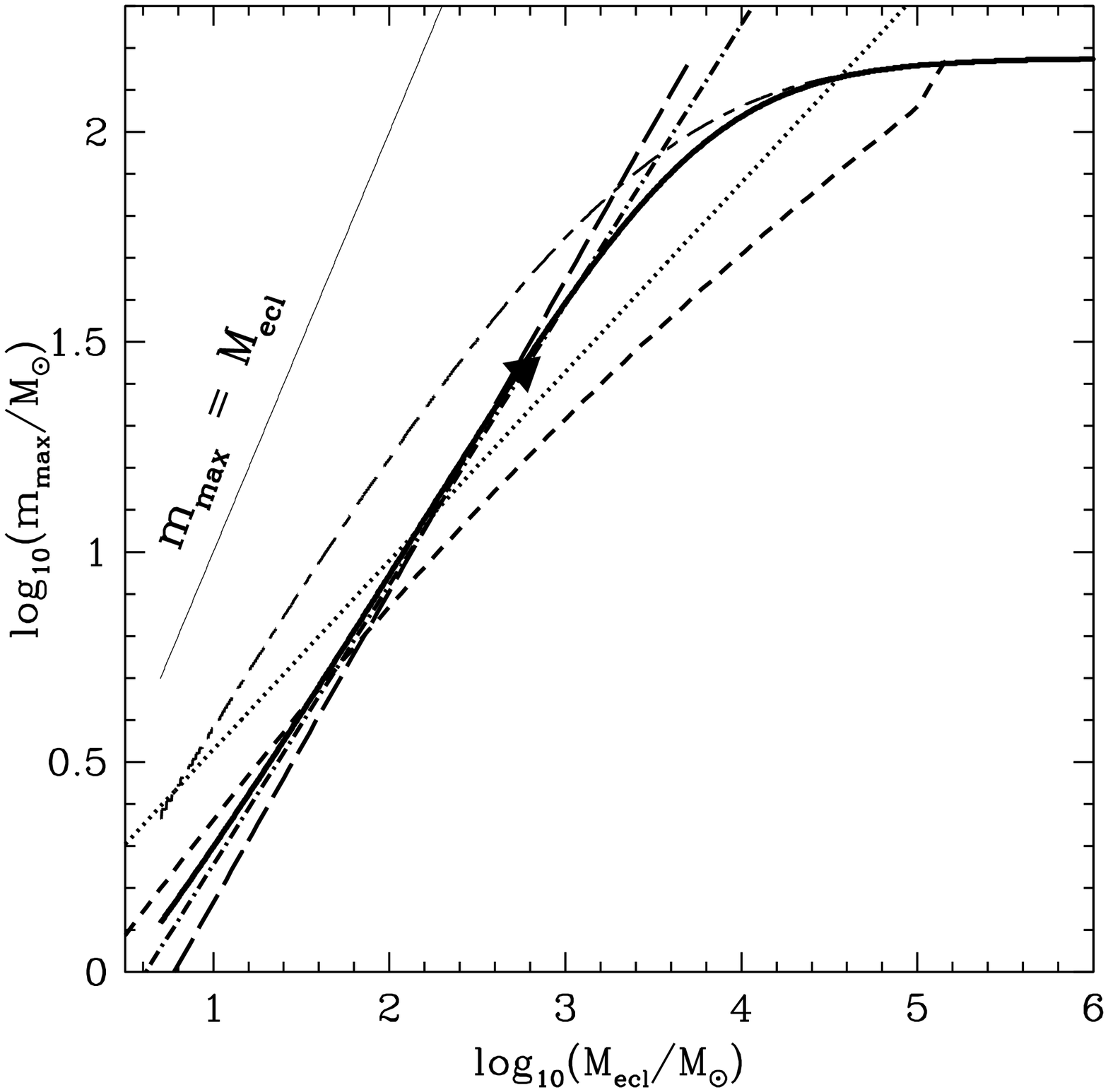}
\vspace*{-2cm}
\caption{The maximal star mass, $m_{\rm max}$, in dependence of the 
  cluster mass, $M_{\rm ecl}$, both in logarithmic units. The
  \citet{La03}-relation (eq.~\ref{eq:La03}) is shown as a {\it dotted
    line}, while the \citet{El83}-relation (eq.~\ref{eq:El83}) is the
  {\it short-dashed line}. The \citet{El00}-relation
  (eq.~\ref{eq:El00}) is the {\it long-dashed line}. The result
  (eq.~\ref{eq:MvMB}) from numerical SPH star-formation simulations
  \citep[][]{BVB04}  is plotted as a {\it short-dashed-dotted line}
  and as a {\it triangle} for a specific model from \citet{BBV03},
  while the {\it long-dashed-short-dashed line} marks the expectation
  values (eq.~\ref{eq:OC05}) from \citet[][]{OC05} and the {\it thick
    solid line} shows the semi-analytical model
  (eq.~\ref{eq:MmaxvsMecl}) of \citet[][]{WK04}  for the standard IMF
  (eq.~\ref{Kroupa-IMF}). The {\it thin solid line} shows the identity
  relation, were a ``cluster'' consists only of one star.
\label{fig:MvsM3}}
\end{center}
\end{figure}

\citet{La82} compared the properties of molecular clouds with the
spatial distribution of the associated stellar population. He found
that more massive and dense clouds favour the formation of massive
stars and fitted the following formula to the observations,
\begin{equation}
\label{eq:La82}
m_{\rm max} = 0.33 M_{\rm cloud}^{0.43}.
\end{equation}
He re-evaluated \citep{La03} this equation with more recent data and
applied it directly to the cluster mass instead of the cloud mass,
\begin{equation}
\label{eq:La03}
m_{\rm max} = 1.2 M_{\rm ecl}^{0.45}.
\end{equation}
This correlation is plotted as a {\it dotted line} in
Figs.~\ref{fig:MvsM3} and \ref{fig:MvsM2}. 

\citet{El83} proposed a model for the origin of bound galactic
clusters {where the luminosity of the stars from a \citet{MS79}
IMF overcomes the binding energy of a molecular cloud core. The
star-formation efficiency then discriminates between bound clusters
and OB associations. He derived an analytical estimate for a relation
between the maximal star mass and the cluster mass from statistical
considerations regarding the appearance of stars with various masses
from the Miller-Scalo-IMF,}
\begin{equation}
\label{eq:El83}
M_{\rm ecl} = \frac{e^{A_{2}}F[A_{1}(\log m_{\rm
    max}-A_{3})]}{1-F[A_{1}(\log m_{\rm max} -C_{2})]},
\end{equation}
with
\begin{equation}
F(x) = (2\pi)^{-1/2} \int_{-\infty}^{x} e^{(-t^{2}/2)} dt,
\end{equation}
and $A_{1} = (2C_{1})^{1/2}$, $A_{2} = \ln 10 (C_{2} + \ln 10 /
4C_{1}) = -1.064$, $A_{3} = C_{2} + \ln 10 / 2C_{1} = 0.065$, $C_{1} =
1.09$ and $C_{2} = -0.99$. In Figs.~\ref{fig:MvsM3} and
\ref{fig:MvsM2} this relation is shown as a {\it short-dashed line}.

On the other hand, using a single power-law IMF with a \citet{Sal55}
slope, \citet{El00} solved eqs.~\ref{eq:mmmc} and \ref{eq:Meclmc},
assuming $m_{\rm max*} = \infty$, with the result,
\begin{equation}
\label{eq:El00}
M_{\rm ecl} = 3 \times 10^{3} \left(\frac{m_{\rm max}}{100
  M_{\odot}}\right)^{1.35} M_{\odot},
\end{equation}
which is shown as a {\it long-dashed line} in Figs.~\ref{fig:MvsM3}
and \ref{fig:MvsM2}. 

\citet{BBV03} and \citet{BVB04} numerically studied star-formation in
clusters using their smooth-particle-hydrodynamics (SPH) code. Here a
turbulent molecular cloud fragments hierarchically into smaller
subunits. When the density of a clump gets higher than a critical
value, it is replaced by a so-called 'sink' particle which only lets
matter in but not out. These sink particles form the final
stellar cluster by interactions and mergings and are called stars at the
end of the simulation. This hierarchical cluster formation scenario
leads to the relation, 
\begin{equation}
\label{eq:MvMB}
m_{\rm max} \propto M_{\rm ecl}^{2/3},
\end{equation}
and is shown as a {\it short-dashed-dotted line} in
Figs.~\ref{fig:MvsM3} and \ref{fig:MvsM2}. There eq.~\ref{eq:MvMB} is
normalised to $m_{\rm max}\,=\,27$ for a cluster of $M_{\rm
  ecl}\,=\,580\,M_{\odot}$ \citep{BBV03,BVB04}. It should be noted
here that these simulations do not include magnetic fields, stellar
feedback {and stellar mergers, all} of which are believed to be of great
importance for star-formation.

\citet{WK04} started with similar assumptions as \citet{El00} but
included a physical upper limit for the stellar mass, $m_{\rm max
  *}\,=\,150\,M_{\odot}$, while solving eqs.~\ref{eq:mmmc} and
\ref{eq:Meclmc}. Due to $m_{\rm max *}$ and using the standard
multi-power law IMF (eq.~\ref{Kroupa-IMF}), the result cannot be
written out analytically but the equations have to be solved
numerically. The result is plotted as a {\it thick solid line} in
Figs.~\ref{fig:MvsM3} and \ref{fig:MvsM2}.

In a broader statistical analysis, \citet{OC05} calculated the
probabilities that the observed upper mass limits in several clusters
and OB associations in the MW, LMC and SMC come from a sample with a
fundamental upper mass limit, $m_{\rm max*}$, or not. They conclude
that an upper mass limit between 120 and 200 $M_{\odot}$ is the most
likely explanation for the observed maximum masses. In oder to do so,
they calculated the expectation value for the maximum mass if a number
$N$ of stars is randomly sampled from an IMF. This yields the
equation,
\begin{equation}
\label{eq:OC05}
\langle m_{\rm max} \rangle = m_{\rm max*} - \int_{0}^{m_{\rm
    max*}}\left[\int_{0}^{M_{\rm ecl}}\xi(m)dm\right]^{N}dM_{\rm ecl}.
\end{equation}
{Integrating this numerically yields the {\it long-dash-short-dashed
  line} in Figs.~\ref{fig:MvsM3} and \ref{fig:MvsM2} assuming $m_{\rm
  max *}\,=\,150\,M_{\odot}$.}

\subsection{Monte-Carlo experiments}
\label{se:MC}
All the above mentioned investigations suggest that the cluster mass
indeed appears to have a limiting influence on the stellar masses within
it. However, it would be undisputed that a stochastic sampling effect
from the IMF must be present when stars form. To investigate the
possible existence of a maximal stellar mass in clusters
statistically and to confirm or rule out if such a relation is purely the
result of the random selection from an IMF, three Monte-Carlo
experiments are conducted:
\begin{itemize}
\item[-] pure random sampling (random sampling),
\item[-] mass-constrained random sampling (constrained sampling),
\item[-] mass-constrained random sampling with sorted adding (sorted sampling).
\end{itemize}
For each, two possibilities are probed: stars are sampled from the IMF
without a maximal mass ($m_{\rm max *} = \infty$\footnote{For practical
reasons of numerical computation $m_{\rm max *}$ is adopted to be
$m_{\rm max *}\,=\,10^{6}\,M_{\odot}$ in the unlimited case.}), or
their masses are limited by $m_{\rm max *}\,=\,150\,M_{\odot}$.

\subsubsection{Random sampling}
\label{se:rasa}
\begin{figure}
\begin{center}
\vspace*{-0.6cm}
\includegraphics[width=8cm]{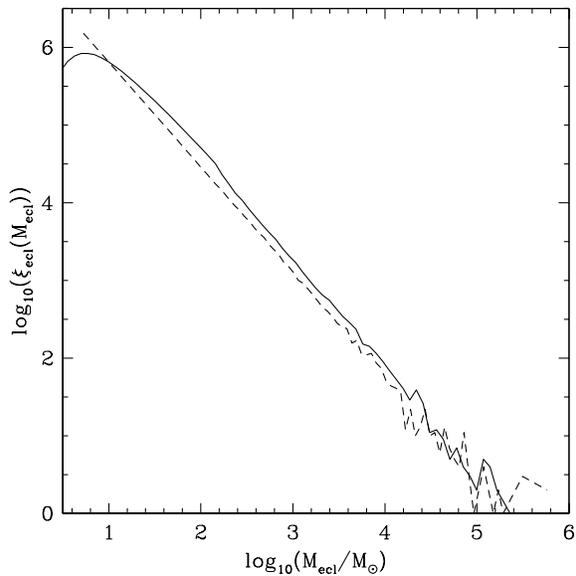}
\vspace*{-2.0cm}
\caption{The embedded cluster mass
  function (ECMF) derived from randomly picking $10^{7}$ clusters
  using eq.~\ref{eq:ECMF} ({\it dashed line}) and as described in
  \S~\ref{se:rasa} ({\it solid line}). The slopes are virtually the
  same ($\beta_{N}\,=\,\beta$). Only for very small cluster masses
  does the {\it solid line} deviate due to the underlying standard IMF
  because small-$N$ clusters can nevertheless have masses
  $>\,10\,M_{\odot}$ if their constituent stars happen to be
  massive. This accounts for the turn down below 10 $M_{\odot}$ and
  the surplus in the range 10-150 $M_{\odot}$.
\label{fig:ECMF}}
\end{center}
\end{figure}

For the random sampling experiment, $2.5 \times 10^{7}$ clusters are
randomly taken from a cluster distribution with a power-law index
$\beta_{N} = 2.35$. {The clusters contain} between 12 stars and $2.7
\times 10^{7}$ stars. The relevant distribution function is the
embedded-cluster star-number function (ECSNF),
\begin{equation}
dN_{\rm ecl} \propto N_{\rm stars}^{\beta_{N}},
\end{equation}
which is the number of clusters containing $N \in [N',N'+dN']$
stars. Each cluster is then filled with $N$ stars randomly from the
standard IMF (eq.~\ref{Kroupa-IMF}) without a mass limit, or by
imposing the physical stellar mass limit, $m\,\le\,150\,M_{\odot}$. 
The stellar masses are then summed to get the cluster mass, $M_{\rm
  ecl}$. Note that such a cluster distribution gives an
embedded cluster mass function (ECMF) that is virtually identical to
eq.~\ref{eq:ECMF} (Fig.~\ref{fig:ECMF}). The resulting distribution of
maximum stellar masses is plotted in {the $m_{\rm max}$, $M_{\rm
    ecl}$ plane} (Fig.~\ref{fig:contourran}) as contour lines, to show
the overall distribution (for a more detailed discussion see
\S~\ref{se:comp}).

As in this method the higher cluster masses are only very scarcely
sampled, a second method is used to evaluate the mean of
maximal masses in more detail. In order to do so the cluster {star
numbers $N = $ 12 to $2.7 \times 10^{7}$ are divided into 10 logarithmically
equally-spaced values.} Each of these numbers is then filled 10000
times with stars form the IMF, keeping only the mass of the most
massive star for each cluster. The mean maximum mass for every bin
then defines,
\begin{equation}
\label{eq:MvsMranl}
^{l}\overline{m}_{\rm max}^{\rm ran} (M_{\rm ecl}),
\end{equation}
 with a limit of $m_{\rm max *}\,=\,150\,M_{\odot}$ and 
\begin{equation}
\label{eq:MvsMranu}
^{u}\overline{m}_{\rm max}^{\rm ran} (M_{\rm ecl}),
\end{equation}
in the unlimited case.

\subsubsection{Constrained sampling}
\label{se:cosa}
In this case $2.5 \times 10^{7}$ clusters are randomly taken from the ECMF
(eq.~\ref{eq:ECMF}) between 5 $M_{\odot}$ (the minimal,
Taurus-Auriga-type, star-forming ``cluster'' counting $\approx$ 15
stars) and $10^{6}\,M_{\odot}$ \citep[an approximate maximum mass for a
single stellar population that consists of one metallicity and
age,][]{WKL04,GLB05} and again with $\beta=2.35$. Note that $\beta_{N}
\approx \beta$ because the ECSNF and the ECMF only differ by a
nearly-constant average stellar mass (Fig.~\ref{fig:ECMF}). Then stars
are taken randomly from the standard IMF and added until they reach or just
surpass the respective cluster mass, $M_{\rm ecl}$. Afterwards the
clusters are searched for their maximum stellar mass (plotted as
contours in Fig.~\ref{fig:contourcon}, see \S~\ref{se:comp} for
discussion).
 
Again the sampling of high cluster masses is very poor. Therefore, the
cluster masses from $5\,M_{\odot}$ to $10^{6}\,M_{\odot}$ are divided
into 10 logarithmically {equally-spaced values. Then each of these 10
cluster masses is filled 10000 times with stars form the IMF
until their combined mass reach or just surpass the cluster mass and
only the maximum star mass is recorded. The average $m_{\rm max}$
values for each of the 10 cluster masses define the relation,}

\begin{equation}
\label{eq:MvsMconl}
^{l}\overline{m}_{\rm max}^{\rm con} (M_{\rm ecl}),
\end{equation}
in the limited case and
\begin{equation}
\label{eq:MvsMconu}
^{u}\overline{m}_{\rm max}^{\rm con} (M_{\rm ecl}),
\end{equation}
in the unlimited case.

\subsubsection{Sorted sampling}
\label{se:sosa}
For the third method again $2.5 \times 10^{7}$ clusters are randomly
taken from the ECMF (eq.~\ref{eq:ECMF}) between 5 $M_{\odot}$ and
$10^{6}\,M_{\odot}$ and with $\beta=2.35$. However, this time the
number $N$ of stars which are to populate the cluster is estimated from
$N\,=\,M_{\rm ecl}$ / $m_{\rm av}$, where $m_{\rm
  av}\,=\,0.36\,M_{\odot}$ is the average stellar mass for the
standard IMF (eq.~\ref{Kroupa-IMF}) between 0.01 $M_{\odot}$ and 150
$M_{\odot}$. These stars are added to give $M_{\rm ecl,ran}$, 
\[M_{\rm ecl,ran} = \sum_{\rm N} m_{i},\]
such that $m_{i} \le m_{i+1}$. If $M_{\rm ran} < M_{\rm ecl}$ in this
first step, an additional number of stars, $\Delta \rm N$, is picked
randomly from the IMF, where $\Delta \rm N$~=~$(M_{\rm ecl} - M_{\rm
  ecl,ran}$)~/~$m_{\rm  av}$. Again these stars are added to obtain an
improved estimate of the desired cluster mass,
\[^{*}M_{\rm ecl,ran} = \sum_{\rm N + \Delta N} m_{i},
\]
again with $m_{i} \le m_{i+1}$. When $^{*}M_{\rm ecl,ran}$ surpasses
$M_{\rm ecl}$, it is checked whether the sum is closer to $M_{\rm
  ecl}$ when the last star {(the most massive one)} is subtracted or not. If
$^{*}M_{\rm ecl,ran}$ is closer to $M_{\rm ecl}$ with the last star
this one is the most massive one for the cluster, otherwise it is the
second last star \footnote{For example, if $M_{\rm
    ecl}\,=\,50\,M_{\odot}$, $m_{N}\,=\,10\,M_{\odot}$ and
  $m_{N-1}\,=\,5\,M_{\odot}$, then for $^{*}M_{\rm
    ecl,ran}\,=\,52\,M_{\odot}$, $m_{\rm max}$ would be 10
  $M_{\odot}$, as 52 is closer to 50 than 42. If $M_{\rm ecl}$ would
  be 45 $M_{\odot}$, $m_{\rm max}$ would be 5 $M_{\odot}$.}. This
procedure is followed and repeated until all clusters from the ECMF
are 'filled'. The contour plots of the most massive star for each
cluster are shown in Fig.~\ref{fig:contoursort} and discussed in
\S~\ref{se:comp}.

Again, for a more detailed analysis, {10 cluster masses are
generated} like in \S~\ref{se:cosa} but filled with stars in the sorted
way described above. The mean over every of the 10 cluster masses then
yields the relation
\begin{equation}
\label{eq:MvsMsortl}
^{l}\overline{m}_{\rm max}^{\rm sort} (M_{\rm ecl}).
\end{equation}

\subsubsection{Comparison of the samplings}
\label{se:comp}
{In order to exemplify the differences between the three sampling
methods the following gedanken experiment may be considered:\\
A sample of 10 stars consists of 9 stars with 5 $M_{\odot}$ and one
with 11 $M_{\odot}$. For the random sampling, this would be a cluster
with $M_{\rm ecl,\,ran}$ = 56 $M_{\odot}$, with $m_{\rm max,\,ran}$ = 11
$M_{\odot}$. If, for the sorted sampling, the aimed-at cluster mass is
50 $M_{\odot}$, the actual cluster mass would be 45 $M_{\odot}$ ($ = 9
\times 5$), because 45 is closer to 50 than 56, and $m_{\rm
  max,\,sort}$ would be 5 $M_{\odot}$. In the case of constrained
sampling the order would be important. If the aimed-at cluster mass is
50 $M_{\odot}$ and the 11 $M_{\odot}$ star is among the first 9 stars,
$M_{\rm ecl,\,con}$ would be 51 ($ = 8 \times 5 + 11$) and $m_{\rm
  max,\,con}$ = 11 $M_{\odot}$. But if the 11 $M_{\odot}$ star is the
tenth star, $M_{\rm ecl,\,con}$ would be 45 $M_{\odot}$ and $m_{\rm
  max,\,con}$ = 5 $M_{\odot}$ as in the sorted-sampling case because,
45, rather than 56, is closer to 50 $M_{\odot}$.\\}

\begin{figure}
\begin{center}
\vspace*{-4.5cm}
\includegraphics[width=9cm]{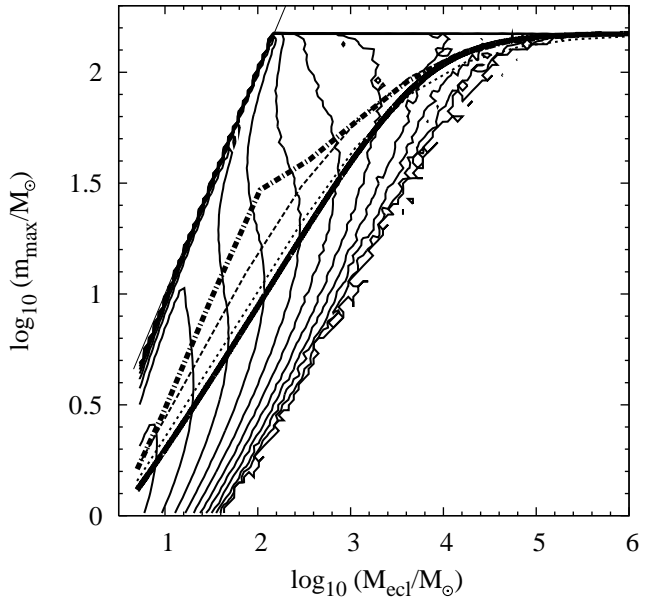}
\caption{Maximal mass of stars versus cluster
  mass. The contour lines show how often a certain combination of
  cluster mass and maximal star mass occurs. The further-out they lie
  (towards the upper right) the smaller is the number of clusters with
  this combination. They are spaced logarithmically such that the
  outer-most one represents a single cluster with a certain mass and
  maximal star ($\log_{10}{N_{\rm ecl}}~=~0$), eg.~$M_{\rm
    ecl}~=~10^{5}\,M_{\odot},\,m_{\rm
    max}~\approx~100\,M_{\odot}$. The inner-most one (near $M_{\rm
    ecl}~=~7\,M_{\odot},\,m_{\rm max}~\approx~{\rm few}
  M_{\odot}$) stands for $\log_{10}{N_{\rm ecl}} = 4.5$ clusters with
  this maximal star mass. The lines in-between are in steps of 0.5
  dex. Here the {\it contour lines} are the result of the Monte-Carlo
  simulations with random sampling. The semi-analytic result
  (eq.~\ref{eq:MmaxvsMecl}), ${}^{l}m_{\rm 
    max}^{\rm ana} (M_{\rm ecl})$, is the {\it thick solid line}. Mean
  values are shown as the {\it thick dash-dotted line} for random
  sampling (${}^{l}\overline{m}_{\rm max}^{\rm ran} (M_{\rm ecl})$,
  eq.~\ref{eq:MvsMranl}), as the {\it thin dashed line} for
  constrained sampling (${}^{l}\overline{m}_{\rm max}^{\rm con} (M_{\rm
    ecl})$, eq.~\ref{eq:MvsMconl}) and as the {\it thin dotted line}
  for sorted sampling (${}^{l}\overline{m}_{\rm max}^{\rm sort} (M_{\rm
    ecl})$, eq.~\ref{eq:MvsMsortl}). The identity relation $m_{\rm 
    max} = M_{\rm ecl}$ is plotted as a {\it thin solid line}.
\label{fig:contourran}}
\end{center}
\end{figure}

\begin{figure}
\begin{center}
\vspace*{-4.5cm}
\includegraphics[width=9cm]{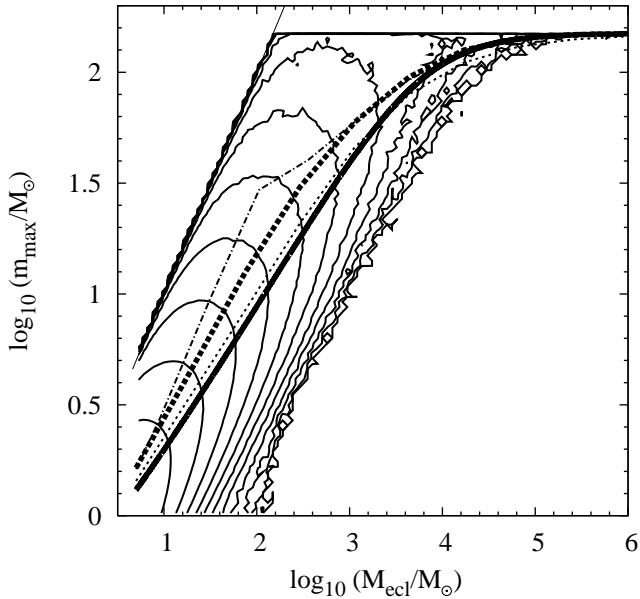}
\caption{Like Fig.~\ref{fig:contourran} but this
  time the {\it contour lines} are the result of the Monte-Carlo
  simulations with mass-constrained sampling, and
  ${}^{l}\overline{m}_{\rm max}^{\rm con} (M_{\rm ecl})$ is shown as a
  {\it thick line}, while ${}^{l}m_{\rm max}^{\rm ran} (M_{\rm ecl})$
  and ${}^{l}\overline{m}_{\rm max}^{\rm sort} (M_{\rm ecl})$ are {\it
    thin}. {$^{l}m_{\rm max}^{\rm ana}$ is the {\it thick solid line}}.
\label{fig:contourcon}}
\end{center}
\end{figure}

\begin{figure}
\begin{center}
\vspace*{-4.5cm}
\includegraphics[width=9cm]{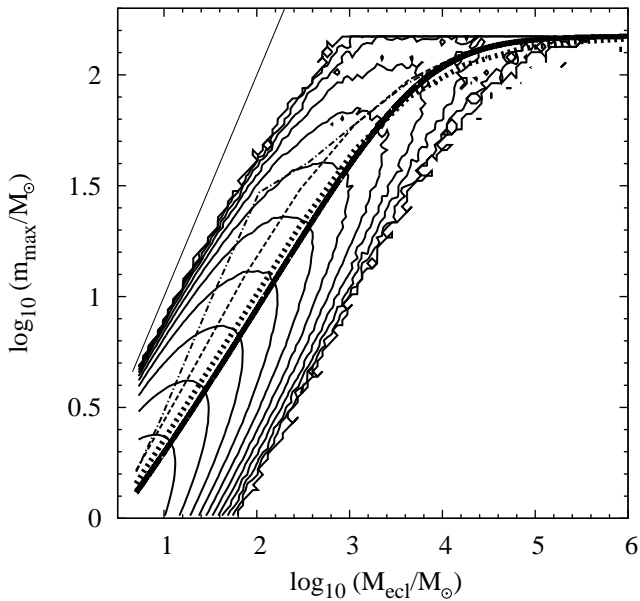}
\caption{Like Fig.~\ref{fig:contourran} and \ref{fig:contourcon} but
  this time the {\it contour lines} are the result of the Monte-Carlo
  simulations with sorted sampling with ${}^{l}\overline{m}_{\rm
    max}^{\rm sort} (M_{\rm ecl})$ as a {\it thick line}. 
\label{fig:contoursort}}
\end{center}
\end{figure}

Figs.~\ref{fig:contourran}, \ref{fig:contourcon} and
\ref{fig:contoursort} plot the contour lines of the most massive stars
of all the simulated clusters for random sampling
(Fig.~\ref{fig:contourran}), constrained sampling
(Fig.~\ref{fig:contourcon}) and sorted sampling
(Fig.~\ref{fig:contoursort}), all with the physical limit $m_{\rm
  max*}\,=\,150\,M_{\odot}$. 

As can be seen from Fig.~\ref{fig:contourran}, the random sample
occupies the whole parameter space between the 
two extremes, which are either that nearly the whole cluster consists
only of low-mass stars, or a single star accounts for the entire
cluster mass ($m_{\rm max}\,=\,M_{\rm ecl}$). Such
one-star-clusters would correspond to freak star-formation events such
as is envisaged for a variable gas-equation-of-state-star-formation
theory \citep{LMK05}. For sorted sampling the clusters 
are shifted more towards smaller stellar masses and never touch this
line. While the constrained sampling lies in between these two extremes
and still populates the parameter space up to the identity relation.


Note that ${}^{l}\overline{m}_{\rm max}^{\rm sort} (M_{\rm ecl})$, shown in
Figs.~\ref{fig:contourran} to \ref{fig:contoursort} ({\it thick dotted
  line} in Fig.~\ref{fig:contoursort}, {\it thin} in
Figs.~\ref{fig:contourran} and \ref{fig:contourcon}), is nearly
identical to the semi-analytical estimate ${}^{l}m_{\rm max}^{\rm ana}
(M_{\rm ecl})$ (eq.~\ref{eq:MmaxvsMecl}, {\it thick solid line} in
Figs.~\ref{fig:contourran} to \ref{fig:contoursort}). The slight
deviations are probably due to the stochastic element in the process
to decide which is the most massive star in the sorted Monte-Carlo
experiment (see \S~\ref{se:sosa}). 

The non-smoothness of the contour lines in the upper right edge of the
figures is  an effect of low-number sampling. Here only the outer-most
contour line is present - indicating that there are only single events
in this region of parameter space obtained with $2.5 \times 10^{7}$
clusters, which have together about $600\,\cdot 10^{6}$ stars. This
number of clusters comprises the computational limit of the available
hardware.

The agreement of the mean value ${}^{l}\overline{m}_{\rm max}^{\rm sort}
(M_{\rm ecl})$ and the semi-analytic result of \citet{WK04},
${}^{l}m_{\rm max}^{\rm ana}(M_{\rm ecl})$, is in principal not
surprising. The method of sorted adding of {stellar} masses in order to get
the cluster mass corresponds to a Monte-Carlo integration of
eqs.~\ref{eq:mmmc} and \ref{eq:Meclmc}. Therefore the result should agree 
with the numerically integrated (semi-analytical) one.

Another difference between the samplings lies in the average mean
stellar masses in the clusters. To determine these, cluster masses from
$5\,M_{\odot}$ to $10^{6}\,M_{\odot}$ are divided in 10
logarithmically equally-spaced values and each is filled 10000 times with stars
from the IMF in the three different ways described in
\S~\ref{se:rasa}, \ref{se:cosa} and \ref{se:sosa}. The mean stellar mass
for each cluster is calculated by
\begin{equation}
\overline{m} = \frac{1}{N} \sum_{N} m_{\rm i},
\end{equation}
where $N$ is the number of stars in each cluster. Then for each
cluster mass, $M_{\rm ecl, j}$ ($j=1...10$), the 'mean of means' is
computed by
\begin{equation}
\label{eq:mofm}
\overline{m}_{\rm j}(M_{\rm ecl, j}) = \frac{1}{N_{\rm ecl}} \sum_{N_{\rm
    ecl}} \overline{m_{\rm i}},
\end{equation}
with $N_{\rm ecl}$ = 10000.

\begin{figure}
\begin{center}
\vspace*{-0.6cm}
\includegraphics[width=8cm]{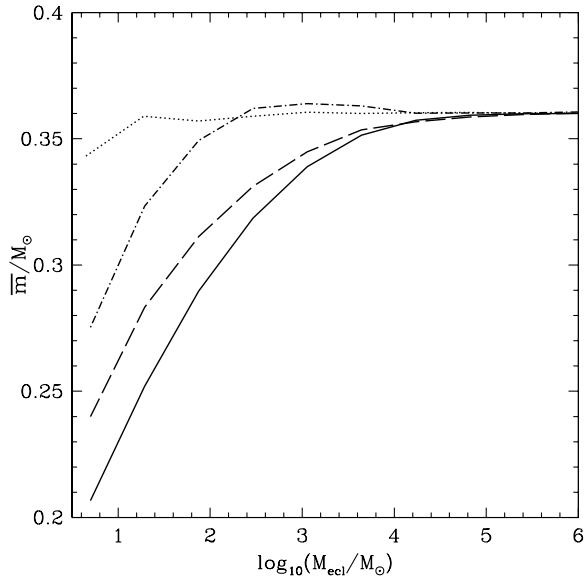}
\vspace*{-2.0cm}
\caption{Average mean stellar mass for the different
  Monte-Carlo samples against cluster mass (eq.~\ref{eq:mofm}). For
  random sampling ({\it dotted line}) the value is constant within the
  numerical noise around the expected value of $0.36\,M_{\odot}$. In
  the case of constrained sampling ({\it dashed-dotted line}) and
  sorted sampling ({\it long-dashed line}) it starts rather low for
  low mass clusters and rises up to expected value of
  $0.36\,M_{\odot}$. The {\it solid line} shows the relation described by
  eq.~\ref{eq:mofmana}. Note that the {\it long-dashed line} lies
  above the {\it solid curve} because the maximal stellar mass in the
  sorted-sampling algorithm is systematically higher than the
  analytical result (Fig.~\ref{fig:MvsM2} to \ref{fig:MvsMage} below)
  for cluster masses below $10^4\,M_{\odot}$.
\label{fig:Mean}}
\end{center}
\end{figure}

The different average mean stellar masses are shown in
Fig.~\ref{fig:Mean}. For the random sampling the mean is always about
$0.36\,M_{\odot}$, as expected for the canonical IMF between 0.01 and
$150\,M_{\odot}$. The other sampling methods have lower means for light 
clusters which rise up to $0.36\,M_{\odot}$ for more massive
ones. This is a result of the limit which low-mass clusters impose on
their stellar content. For random sampling only the number of stars
determines the cluster mass, in contrast to reality where the natal
cloud-mass and the star-formation-efficiency rule the final cluster mass.

Also shown in Fig.~\ref{fig:Mean} is the average mean stellar mass for
the analytical model (eq.~\ref{eq:MmaxvsMecl}), which is given by
\begin{equation}
\label{eq:mofmana}
\overline{m} = \frac{\int_{m_{\rm low}}^{m_{\rm max}(M_{\rm ecl})}m\xi
  dm}{\int_{m_{\rm low}}^{m_{\rm max}(M_{\rm ecl})}\xi dm},
\end{equation}
where $\xi(m)$ is eq.~\ref{Kroupa-IMF}.

\subsubsection{Comparison with observational data}

\begin{figure}
\begin{center}
\vspace*{-0.6cm}
\includegraphics[width=8cm]{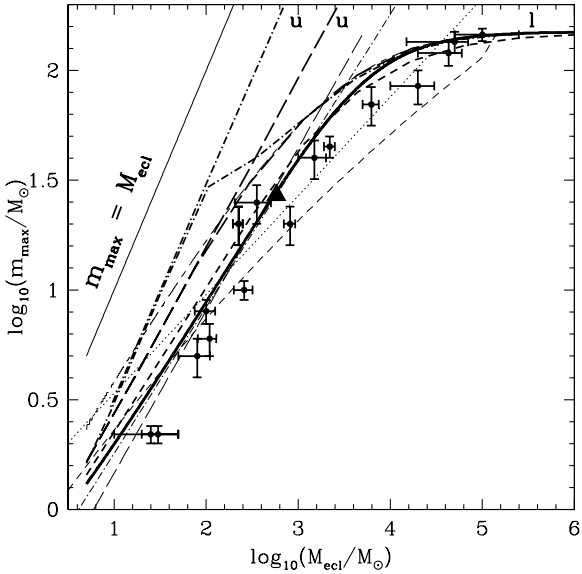}
\vspace*{-2cm}
\caption{The {\it thick solid line}
  shows the dependence of the maximal star mass on the cluster mass
  for $\alpha_{3} = 2.35$  from the semi-analytical model
  (${}^{l}m_{\rm max}^{\rm ana} (M_{\rm ecl})$,
  eq.~\ref{eq:MmaxvsMecl}). The {\it thick short-dashed line}
  shows the mean maximum stellar mass for sorted sampling
  (${}^{l}\overline{m}_{\rm max}^{\rm sort}(M_{\rm ecl})$, see
  \S~\ref{se:sosa}). The {\it long-dashed lines} are mass-constrained
  random-sampling (${}^{l,u}\overline{m}_{\rm max}^{\rm con}(M_{\rm
    ecl})$, see \S~\ref{se:cosa}) results with an upper mass limit of
  $10^{6}\,M_{\odot}$ ({\it straight line}) and $150\,M_{\odot}$ ({\it
    curved line}). Pure random sampling models
  (${}^{l,u}\overline{m}_{\rm max}^{\rm ran}(M_{\rm ecl})$, see
  \S~\ref{se:rasa}) are printed as {\it dot-dashed lines}. The {\it
    curved} one is sampled to $m_{\rm max}\,=\,150\,M_{\odot}$ while
  the {\it straight} one assumes $m_{\rm max
    *}\,=\,10^{6}\,M_{\odot}$. The {\it thin solid line} shows the
  identity relation, were a ``cluster'' consists only of one star. The
  {\it dots} with error bars are observed clusters (see
  Tab.~\ref{tab:clusters}), while the {\it triangle} is a result from
  a star-formation simulation (eq.~\ref{eq:MvMB}) with an SPH code
  \citep[][]{BBV03}. A previous study of massive stars in clusters
  (eq.~\ref{eq:La82} and \ref{eq:La03}) by \citet[][]{La82,La03} is
  shown as a {\it thin dotted line}, while an analytic estimate
  (eq.~\ref{eq:El83}) by \citet[][]{El83} as a {\it thin short-dashed
    line} and eq.~\ref{eq:El00} by \citet[][]{El00} as a 
  {\it thin long-dashed line}. The {\it thin long-dashed-short-dashed line}
  marks the expectation values (eq.~\ref{eq:OC05}) from \citet[][]{OC05}.
\label{fig:MvsM2}}
\end{center}
\end{figure}

\begin{figure}
\begin{center}
\vspace*{-0.6cm}
\includegraphics[width=8cm]{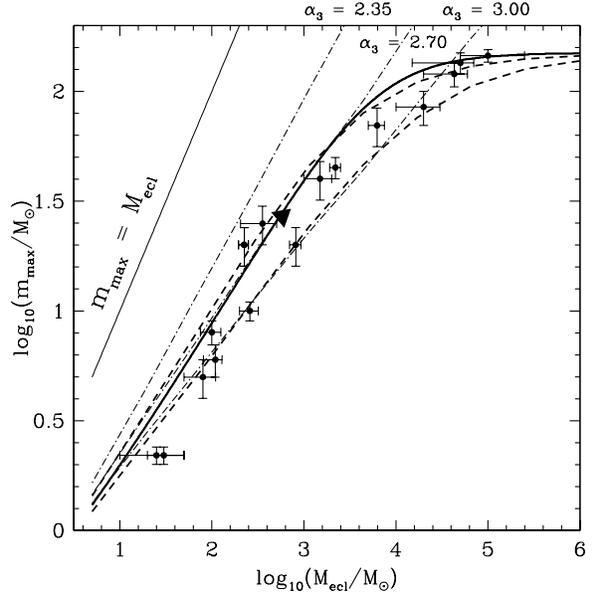}
\vspace*{-2cm}
\caption{The {\it thick solid line} shows
  ${}^{l}m_{\rm max}^{\rm ana} (M_{\rm ecl})$ for $\alpha_{3} = 2.35$
  (eq.~\ref{eq:MmaxvsMecl}). The {\it thick dashed lines} are
  $^{l}\overline{m}_{\rm max}^{\rm sort} (M_{\rm ecl})$ for
  $\alpha_{3}\,=\,2.35$ (upper line) and 2.70 (lower line). The {\it 
    dot-dashed lines} are constrained sampling Monte-Carlo results,
  ${}^{u}\overline{m}_{\rm max}^{\rm con}(M_{\rm ecl})$, with three
  different slopes ($\alpha_{3}\,=\,2.35,\,2.70~{\rm and}~3.00$
  above $1\,M_{\odot}$) of the input stellar IMF. The {\it thin solid
    line} shows the identity relation. The {\it dots} with error bars
  are observed clusters (Tab.~\ref{tab:clusters}), while the {\it
    triangle} is a result from the star-formation simulation with an
  SPH code \citep{BBV03}. 
\label{fig:MvsM}}
\end{center}
\end{figure}

Table~\ref{tab:clusters} shows a non-exhaustive compilation of cluster
masses and upper stellar masses for a number of MW and LMC clusters
(see appendix~\ref{appendixA} for more details). These data are shown
as dots with error-bars in Figs.~\ref{fig:MvsM2},~\ref{fig:MvsM} and
\ref{fig:MvsMage}. The result of a cluster formation simulation by
\citet{BBV03} is shown as a large triangle.

\begin{table}
\centering
\caption{\label{tab:clusters} Empirical cluster masses, maximal star masses
  within these clusters and cluster ages from the literature.}
\vspace*{0.5cm}
\begin{tabular}{cccccc}
\hline
Designation&$M_{\rm ecl}$&$m_{\rm max obs}$&age&Source\\
&[$M_{\odot}$]&[$M_{\odot}$]&[Myr]&\\
\hline
Tau-Aur&25 $\pm$ 15&2.2 $\pm$ 0.2& 1-2&(1)\\
Ser SVS2&30 $\pm$ 15&2.2 $\pm$ 0.2& 2&(2)\\
NGC1333&80 $\pm$ 30&5 $\pm$ 1.0& 1-3&(3)\\
$\rho$ Oph&100 $\pm$ 50&8 $\pm$ 1.0& 0.1-1&(4)\\
IC348&109 $\pm$ 20&6 $\pm$ 1.0& 1.3&(5)\\
NGC2024&225 $\pm$ 30&20 $\pm$ 4& 0.5&(6)\\
$\sigma$ Ori&225 $\pm$ 30&20 $\pm$ 4& 2.5&(6)\\
Mon R2&259 $\pm$ 60&10 $\pm$ 1& 0-3&(7)\\
NGC2264&355 $\pm$ 50&25 $\pm$ 5& 3.1&(8)\\
NGC6530&815 $\pm$ 115&20 $\pm$ 4& 2.3&(9)\\
Ber 86&1500 $\pm$ 500&40 $\pm$ 8& 2-3&(10)\\
M42&2200 $\pm$ 300&45 $\pm$ 5& $<$1&(11)\\
NGC2244&6240 $\pm$ 124&70 $\pm$ 14& 1.9&(12)\\
NGC6611&2$\cdot 10^{4}$ $\pm$ 10000&85 $\pm$ 15&1.3 $\pm$ 0.3&(13)\\
Tr 14/16&4.3$\cdot 10^{4}$ {\scriptsize -2.3$\cdot 10^{4}$/+2$\cdot 10^{4}$}&120 $\pm$ 15& $<$3&(14)\\
Arches& 5$\cdot 10^{4}$ {\scriptsize -3.5$\cdot 10^{4}$/+2$\cdot 10^{4}$}&135 $\pm$ 15& 2.5&(15)\\
R136&1$\cdot 10^{5}$ {\scriptsize -5$\cdot 10^{4}$/+1.5$\cdot 10^{5}$}&145 $\pm$ 10& 1-2&(16)\\
\hline
Simulation&580&27& 0.456 &(17)\\
\hline
\end{tabular}

$^{1}$ \citet{BLH02}, $^{2}$ \citet{KOB04}, $^{3}$ \citet{AC03,GFT02},
$^{4}$ \citet{WLY89,La03}, $^{5}$ \citet{PZ01,LL03}, $^{6}$ \citet{SWW04},
$^{7}$ \citet{CMD97}, $^{8}$ \citet{SBC04}, $^{9}$
\citet{PDM04,DFM04}, $^{10}$ \citet{MJD95,VRC99}, $^{11}$
\citet{HH98,HSC98}, $^{12}$ \citet{MJD95,PS02}, $^{13}$ \citet{BSB05},
$^{14}$ \citet{MJ93,MJD95}, $^{15}$ \citet{FNG02}, $^{16}$
\citet{MH98,SMB99}, $^{17}$ \citet{BBV03} 
\end{table}

Fig.~\ref{fig:MvsM2} compares the mean $m_{\rm max}$ values of the
Monte-Carlo experiments together with the observations from
Tab.~\ref{tab:clusters}, the semi-analytical result of
\citet{WK04} and the different results of the previous' studies
(\S~\ref{se:PS}). The ``unlimited'' mean values are marked with a
``u'', while the limited ones ($m_{\rm max *}\,=\,150\,M_{\odot}$)
with an ``l''. The mean values of the ``u'' cases are all clearly
distinct from the observations and not regarded further, while for the
``l'' cases the mean values are in reasonable agreement with the
observations. But especially ${}^{l,u}\overline{m}_{\rm max}^{\rm ran}
(M_{\rm ecl})$ and ${}^{l,u}\overline{m}_{\rm max}^{\rm con} (M_{\rm
  ecl})$ lie above the observations, while ${}^{l}\overline{m}_{\rm
  max}^{\rm sort} (M_{\rm ecl})$ and ${}^{l}m_{\rm max}^{\rm ana}
(M_{\rm ecl})$ fit the observations rather well.

The observational analysis (eq.~\ref{eq:La82} and \ref{eq:La03}) by
\citet[][]{La82,La03} and the analytical result (eq.~\ref{eq:El83}) of
\citet[][]{El83} have a shallower slope and underestimate the observed
stellar masses for $M_{\rm ecl}\,>\,100\,M_{\odot}$. The analytical
result (eq.~\ref{eq:El00}) of \citet[][]{El00} and the
star-formation simulation (eq.~\ref{eq:MvMB}) by
\citet[][]{BBV03,BVB04} fit much better but ignore the observed upper
mass limit for stars near $150\,M_{\odot}$
\citep{WK04,Fi05,OC05}. Taken this into account leads to the
semi-analytical (eq.~\ref{eq:MmaxvsMecl}) relation. The 
\citet{OC05} expectation values for $m_{\rm max}$ (eq.~\ref{eq:OC05})
follow quite well the constrained sampling result, but in doing so they
over-predict $m_{\rm max}$ in comparison with the observations.

In Fig.~\ref{fig:MvsM} only ${}^{u}\overline{m}_{\rm max}^{\rm con}
(M_{\rm ecl})$  and ${}^{l}\overline{m}_{\rm max}^{\rm sort} (M_{\rm
  ecl})$ are shown but for different slopes of the IMF above
$1\,M_{\odot}$. For ${}^{u}\overline{m}_{\rm max}^{\rm con} (M_{\rm
  ecl})$ $\alpha_{3}$ = 2.35, 2.70 and 3.00 is plotted, while for
${}^{l}\overline{m}_{\rm max}^{\rm sort} (M_{\rm ecl})$ the relations
are plotted for $\alpha_{3}$ = 2.35 and 2.70. In the case of
${}^{u}\overline{m}_{\rm max}^{\rm con} (M_{\rm ecl})$ we can see that
the steeper IMF slopes fit the observations much better. Therefore, if
\citet{SR91} are right about their notion that unresolved binaries
would mask a steep underlying IMF, constrained or even random sampling
cannot be ruled out. But it should be noted here, that the
\citet{SR91} examination was only carried out for stars with masses
between 2 and 14 $M_{\odot}$. Therefore, the O-star regime has not yet
been explored in this respect.

\subsection{Ageing of the stars}
\label{subsec:age}
To see if stellar evolution together with constrained sampling can
mimic the effect of the sorted sampling, stars with $m \le
50\,M_{\odot}$ are evolved with the single stellar evolution (SSE)
package from \citet{HPT00}, while for more-massive stars fitting
formulae derived from {stellar evolution models computed
  by} \citet{SSM92} are used (see 
Appendix~\ref{appendixB} for for the detailed fitting functions). For
this purpose $1\times10^{7}$ clusters are chosen from an ECMF with $\beta
= 2.35$ and then aged for 1, 2, 2.5, 3 and 3.5 Myr. The evolved
stellar masses are added after excluding neutron stars and black holes
to give $M_{\rm ecl}$ and then searched for the most massive star in
each cluster. 

\begin{figure}
\begin{center}
\vspace*{-0.6cm}
\includegraphics[width=8cm]{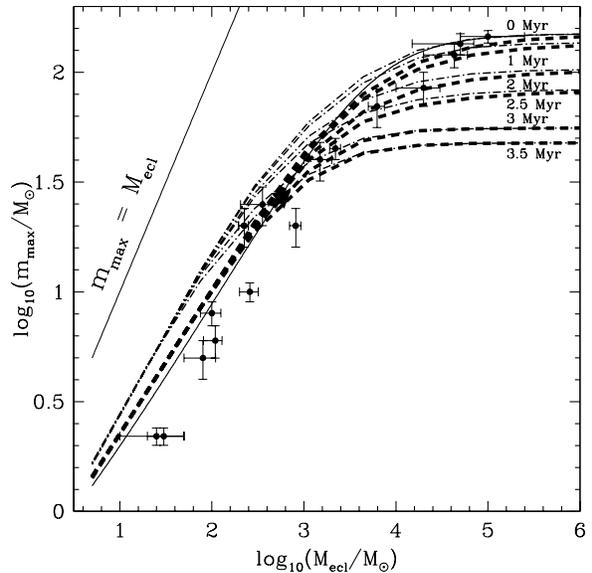}
\vspace*{-2cm}
\caption{As Fig.~\ref{fig:MvsM2} but the mean
  curves include ageing by 1, 2, 2.5, 3 and 3.5 Myr. The stars in the
  Monte-Carlo-simulations are subject to stellar evolution according 
  to the SSE package by \citet{HPT00} and our own extensions for stars
  $\ge\,50\,M_{\odot}$ which includes not only finite life-times
  but also stellar mass-loss. The {\it thick dashed lines} are 
  for clusters which are constructed using sorted sampling, while the {\it
    dot-dashed lines} are for constrained sampling. Note that 
  ${}^{l}\overline{m}_{\rm max}^{\rm con} (M_{\rm ecl}) = {}^{l}\overline{m}_{\rm 
    max}^{\rm sort} (M_{\rm ecl})$ for ages $\protect\simgreat$ 3 Myr
  and $M_{\rm ecl}\,\ge\,10^{3}\,M_{\odot}$.
\label{fig:MvsMage}
}
\end{center}
\end{figure}

In Fig.~\ref{fig:MvsMage} the effect of this ageing is shown. Within
the first 2.5 Myr the mean values of the sorted sampling and the
constrained sampling algorithms are clearly distinct. The lines change  
due to mass loss of the heavy stars, which amounts to about 40 to 50
\% of the initial stellar mass, according to the models used (see
Appendix~\ref{appendixC} for a comparison of different models). After
2.5 Myr the massive stars start to explode as supernovae and the
constrained sampled clusters (${}^{l}\overline{m}^{\rm con}_{\rm max}(M_{\rm
  ecl})$) move closer to the sorted-sampled ones
(${}^{l}\overline{m}^{\rm sort}_{\rm max}(M_{\rm ecl})$) for $M_{\rm
  ecl}\,\ge\,10^{3}\,M_{\odot}$. After 1 Myr and for clusters with $M_{\rm
  ecl}\,\ge\,10^{3}\,M_{\odot}$, ageing shifts ${}^{l}\overline{m}^{\rm
  con}_{\rm max}(M_{\rm ecl})$ closer to the observations, making a
distinction between constrained and sorted sampling not possible. 
Nevertheless, the {observational data for $M_{\rm ecl} <
10^{3}\,M_{\odot}$} agree much better with sorted than with constrained
sampling for all ages.

It must be noted here that the observed stellar masses ($m_{\rm max
  obs}$) in Tab.~\ref{tab:clusters} are a mixture of present-day (PD)
masses and zero-age main-sequence (ZAMS) masses. For stars below
roughly 50 $M_{\odot}$ this is not critical, as for them mass-loss is
not dominant during the first 3 Myr. But the three most-massive clusters
have ZAMS maximal stellar masses and can therefore not be compared
with the aged tracks in Fig.~\ref{fig:MvsMage}. Our previous
(\S~\ref{se:MC}) comparison of the data with zero-age main-sequence
isochrones is thus justified.

\subsection{Statistical analysis}
\label{subsec:random}
In order to evaluate the statistical significance of the differences
between the Monte-Carlo simulations and the observations, two
statistical tests are applied. The statistical tests are only
applied to the zero-age Monte-Carlo samples, as the PD and ZAMS masses
for low-mass stars do not differ substantially for such young clusters
and all the massive stars ($m\,\ge\,50\,M_{\odot}$) in the
observational sample are always ZAMS masses. While for ages $\le$ 2.5
Myr {different} stellar models agree rather reasonably on the
properties and 
evolution of massive stars (see Appendix~\ref{appendixC} for a
substantial comparison of some models) the theoretical foundation of
the formation of such massive stars is still very weak \citep[for a
detailed discussion of massive stars, see][]{KW05}.

\begin{figure}
\begin{center}
\includegraphics[width=8cm]{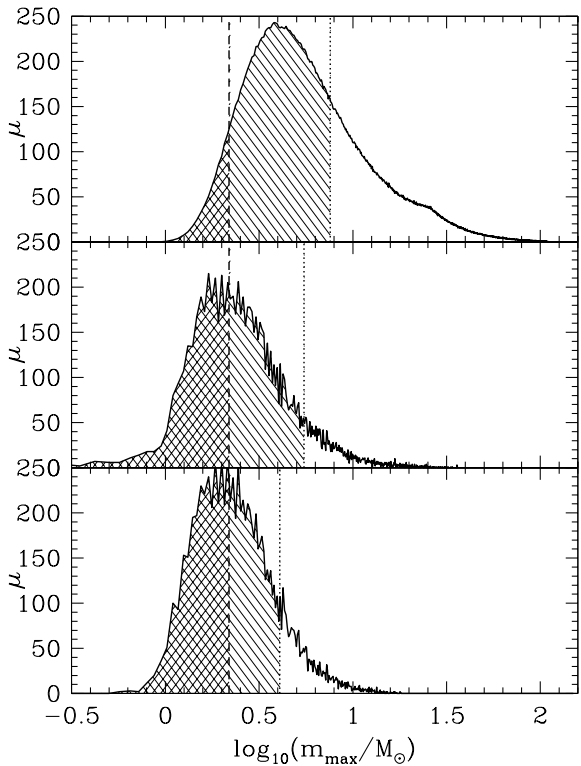}
\caption{{\it All panels:} Vertical slice through 
  Figs.~\ref{fig:contourran}, \ref{fig:contourcon} and
  \ref{fig:contoursort} at $M_{\rm ecl}\,=\,25\,M_{\odot}$
  (Taurus-Auriga). The vertical axis plots the number of clusters
  $\mu$ which have the maximal mass $m_{\rm max}$ in a logarithmic bin
  $\Delta \log_{10}m_{\rm max}\,=\,0.01044$. {\it Top panel:}
  Histogram of the number of clusters for the random sampling models
  (\S~\ref{se:rasa}). The vertical {\it dashed line} marks the
  observed upper mass in Taurus-Auriga, while the {\it dotted line} is
  the mean (or expectation) value of the Monte-Carlo experiment,
  ${}^{l}\overline{m}_{\rm max}^{\rm ran}(M_{\rm ecl})$. The probability
  of observing an $m_{\rm max}$ as small as or smaller than $m_{\rm
    max obs}$ is $P(m_{\rm max}\,\le\,m_{\rm max obs})$ which is the
  area under the curve left of the {\it dashed line} divided by the
  area under the curve left of the {\it dotted line}. {\it Middle
    panel:} Like {\it top panel} but for the constrained-sampling
  Monte-Carlo experiment (\S~\ref{se:cosa}). {\it Bottom panel:} As
  {\it top panel} but in this case for sorted sampling
  (\S~\ref{se:sosa}).{{\it All panels:}  Note that the areas to the
  left and right of the mean value do not appear equal because the
  binning was done linearly but is plotted on a logarithmic scale.}
\label{fig:Int1}
}
\end{center}
\end{figure}

\begin{figure}
\begin{center}
\includegraphics[width=8cm]{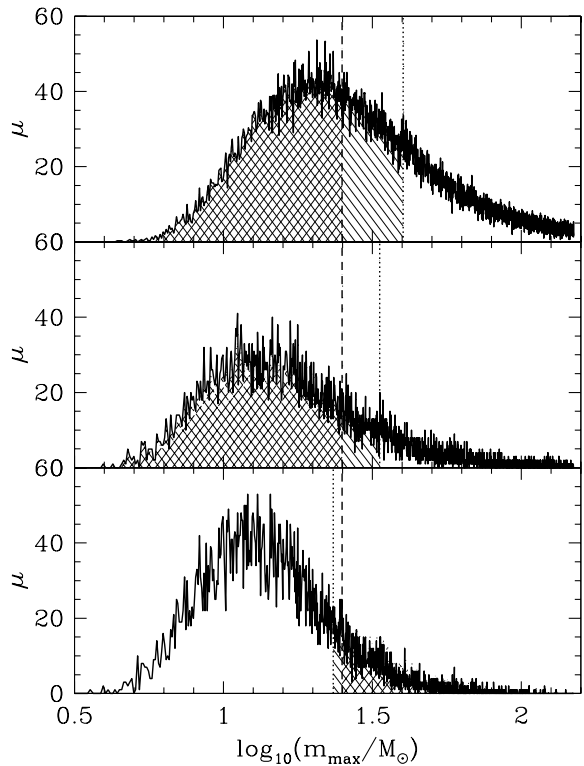}
\caption{All three panels like Fig.~\ref{fig:Int1}
  but in this case for a cluster mass of 355 $M_{\odot}$
  (NGC2264). For the sorted sampling ({\it bottom panel}) the mean is 
  lower than the observed value of $m_{\rm max}$. Therefore the
  probability is calculated by dividing the area to the right of the
  observed value ({\it dashed line}) through the area to the right of
  the expectation value ({\it dotted  line}).
\label{fig:Int2}
}
\end{center}
\end{figure}

The probabilities that the observed masses of the
most-massive stars, $m_{\rm max obs}$, in the clusters are drawn from
the three different Monte-Carlo samples are calculated. In order to do
so, the distribution of $m_{\rm max}$ around each observed
{cluster} mass, $\mu (m_{\rm max}:M_{\rm ecl})$, is examined for the three
samplings, whereby we only use those that have a maximal stellar mass,
$m_{\rm max*}\,=\,150\,M_{\odot}$. If the mean value,
$\overline{m}_{\rm max}(M_{\rm ecl})$, of the Monte-Carlo
distribution, $\mu(m_{\rm max})$, for a cluster mass is higher than
the observed maximal mass, $m_{\rm max obs}$, then the distribution is
integrated from {the lower limit, $m_{\rm low}$, to $m_{\rm max
    obs}$ and divided by the integral from $m_{\rm low}$ to
${}^{l}\overline{m}_{\rm max}^{\rm ran,\,con,\,sort}$,
\begin{equation}
\label{eq:Probhi}
P(m_{\rm max} \le m_{\rm max obs}) = \frac{\int_{m_{\rm low}}^{m_{\rm max obs}}\mu\,
  dm}{\int_{m_{\rm low}}^{{}^{l}\overline{m}_{\rm max}^{\rm
      ran,\,con,\,sort}}\mu\,dm}.
\end{equation}
This is the probability of observing a maximum mass $m_{\rm max}$ as
small as or smaller than $m_{\rm max obs}$. 

If, on the other hand, the mean value is smaller than $m_{\rm max
  obs}$, the integral is taken
from $m_{\rm max obs}$ to the upper mass limit, $m_{\rm max *}$, and
is divided by the integral from the mean value,
${}^{l}\overline{m}_{\rm max}^{\rm ran,\,con,\,sort}$, to $m_{\rm max *}$,
\begin{equation}
\label{eq:Problow}
P(m_{\rm max} \ge m_{\rm max obs}) = \frac{\int_{m_{\rm max obs}}^{m_{\rm max *}}\mu\,
  dm}{\int_{{}^{l}\overline{m}_{\rm max}^{\rm ran,\,con,\,sort}}^{m_{\rm
      max *}}\mu\,dm}.
\end{equation}
This is the probability of observing an $m_{\rm max}$ as large as
or larger than $m_{\rm max obs}$.} Together, eqs.~\ref{eq:Probhi} and
\ref{eq:Problow} are the probability of observing an $m_{\rm max obs}$
such that $|m_{\rm max obs}\,-\,\overline{m}_{\rm max}(M_{\rm
  ecl})|\,\ge\,|m_{\rm max}\,-\,\overline{m}_{\rm max}(M_{\rm
  ecl})|$. Figs.~\ref{fig:Int1} and \ref{fig:Int2} show
a schematic view of this integration process by means of two
examples (Taurus-Auriga and NGC2264) for all three cases of
sampling. 

The resulting combined probabilities ($\Pi$ in
Tab.~\ref{tab:props}, acquired by multiplying the individual
probabilities) for all clusters are $10^{-9}$ for random
sampling, $10^{-6}$ for constrained sampling and $10^{-4}$ for
sorted sampling, with the highest probability thus being attained for
sorted sampling. Tab.~\ref{tab:props} shows the individual
probabilities $P$ for each $m_{\rm max obs}$.

\begin{table}
\centering
\caption[Probabilities for the Monte-Carlo samples]{\label{tab:props}
  Probabilities that the $m_{\rm max obs}$ are from one of the three
  Monte-Carlo samples.} 
\vspace*{0.5cm}
\begin{tabular}{cccc}
\hline
Cluster&random&constrained&sorted\\
&sampling&sampling&sampling\\
&\%&\%&\%\\
\hline
Tau-Aur     &25.0&36.2&45.4\\
Ser SVS2    &12.5&33.1&41.5\\
NGC1333     &24.7&29.4&41.1\\
$\rho$ Oph  &31.5&60.6&81.7\\
IC348       &\phantom{0}9.2&38.2&52.5\\\
NGC2024     &69.1&86.0&76.4\\
$\sigma$ Ori&69.1&86.0&76.4\\
Mon R2      &25.8&27.1&37.5\\
NGC2264     &57.6&81.9&84.7\\
NGC6530     &27.2&32.4&44.3\\
Ber 86      &39.3&53.9&76.3\\
M42         &43.9&53.6&75.5\\
NGC2244     &34.1&40.4&68.0\\
NGC6611     &\phantom{0}6.4&\phantom{0}5.6&32.2\\
Tr 14/16    &42.4&44.8&96.6\\
Arches      &75.0&88.2&67.4\\
R136        &65.7&82.3&42.2\\
\hline
$\Pi$&$3.37 \times 10^{-9}$&$9.22 \times 10^{-7}$&$9.18 \times
10^{-5}$\\
\hline
\end{tabular}
\end{table}


{To obtain a completely different statistic on the correspondence
between data and theory, we also apply} a Wilcoxon signed-rank test
\citep{BJ77} to 
determine how significant the differences between the data and the
${}^{l}\overline{m}_{\rm max}^{\rm ran,\,con,\,sort}$ relations are. For
this test the differences of the data points and the mean values are
calculated,
\begin{equation}
\Delta m_{\rm j} = m_{\rm max obs,\,j} - {}^{l}\overline{m}_{\rm max,\,j}^{\rm
  ran,\,con,\,sort}\,({\rm for~a~given}\,M_{\rm ecl,\,j}),
\end{equation}
and then ranked according to their absolute value. Afterwards, only the
positive-signed ranks are added and this sum is then cross-checked
with tabulated values \citep{BJ77} in order to get the probability $P$
that the null hypothesis (data points and the ${}^{l}\overline{m}_{\rm
  max}^{\rm ran,\,con,\,sort}$ relations are the same within the
uncertainties) is true. In the case of random sampling
(\S~\ref{se:rasa}) and constrained sampling
(\S~\ref{se:cosa}) {$P$ = 0.00014, while $P$ = 0.0458 for sorted sampling
(\S~\ref{se:sosa}).} 

Thus, both tests {taken together} suggest strongly that
sorted-sampling best describes the empirical data. The physical
interpretation of this result is discussed in \S~\ref{sec:concs}.

\section{The Monte-Carlo simulations of the IGIMF}
\label{sec:montec}
We have thus seen that the observational data strongly favour a
particular $m_{\rm max}(M_{\rm ecl})$ relation, namely the
${}^{l}\overline{m}_{\rm max}^{\rm sort}\,\approx\,^{l}{m}_{\rm
 max}^{\rm ana}$ relation. This has profound implications for
composite stellar populations.

\begin{figure}
\begin{center}
\vspace*{-0.6cm}
\includegraphics[width=8cm]{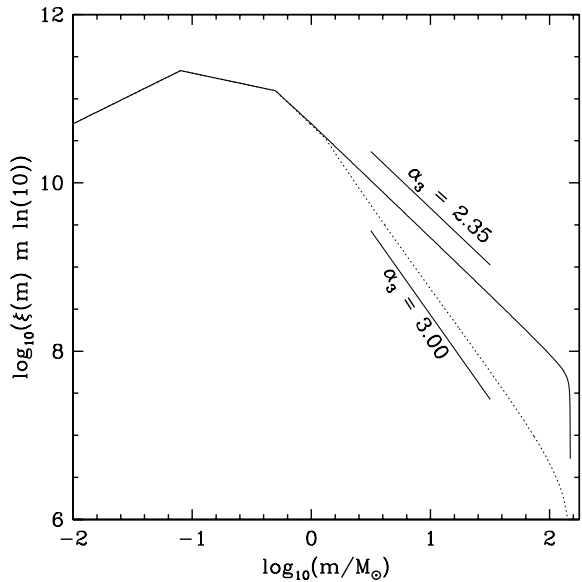}
\vspace*{-2cm}
\caption{{\it Solid line}: Canonical
  stellar IMF, $\xi(m)$, in logarithmic units and given by the
  standard three-part  power-law form (eq.~\ref{Kroupa-IMF}), which has
  $\alpha_{3}=2.35$ for $m>0.5\,M_\odot$. {\it Dotted line}: semi-analytical
  $\xi_{\rm IGIMF}(m)$ for $\beta=2.35$. The IMFs are scaled to have
  the same number of objects in the mass interval $0.01-1.0\,M_\odot$.
  Note the turn down near $m_{\rm max*}=150\,M_\odot$ which comes from
  taking the fundamental upper mass limit explicitly into account
  \citep{WK04}. Two lines with slopes $\alpha_{\rm line}=2.35$ and
  $\alpha_{\rm line}=3.00$ are indicated.
\label{fig:mc1}}
\end{center}
\end{figure}

Fig.~\ref{fig:mc1} shows the result of the 
semi-analytic approach by \citet{KW03} assuming $\beta=2.35$ for the
ECMF.  The stellar IMF in each cluster has, in all cases, the standard
or canonical three-part power-law form (eq.~\ref{Kroupa-IMF}). For the
minimum ``cluster'' mass, $M_{\rm ecl,min}\,=\,5\,M_\odot$ (a dozen stars),
and for the maximal cluster mass, $M_{\rm ecl,max}=10^6\,M_\odot$, are
used. The power-law index $\alpha_{\rm IGIMF}$ of the resulting
semi-analytical IGIMF is well approximated by $\alpha_{\rm
  IGIMF}=3.00$ for $m\simgreat 1\,M_\odot$. In \citet{KW03} we already
noted that this {is probably} the reason why the Galactic-field-IMF deduced
by \citet{Sc86} ($\alpha_{3}\,\approx\,2.7$) is steeper than the
canonical stellar IMF.

\begin{figure}
\begin{center}
\includegraphics[width=8cm]{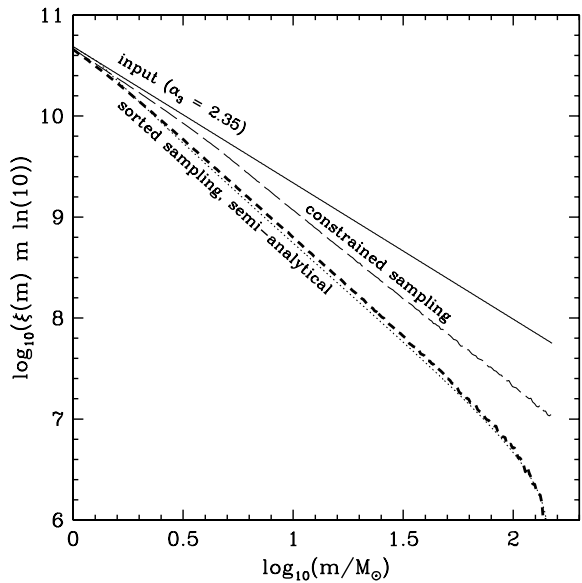}
\vspace*{-2cm}
\caption{{\it Solid
    line}: Standard stellar IMF with $\alpha_{3}=2.35$ for
  $m>1\,M_\odot$ (same as in Fig.~\ref{fig:mc1}). {\it Dotted line}:
  IGIMF resulting from the semi-analytical approach with $\beta=2.35$
  (as in Fig.~\ref{fig:mc1}). {\it Short dashed line}: IGIMF obtained
  from sorted sampling. {\it Long dashed line}: IGIMF produced by
  constrained sampling of stars. As the IMF below $1\,M_{\odot}$ does
  not change, only the region above $1\,M_{\odot}$ is plotted here.
\label{fig:mc2}}
\end{center}
\end{figure}

We now apply the Monte-Carlo experiments introduced above to the
computation of the IGIMF (eq.~\ref{eq:IGIMF}). The resulting IGIMF is
constructed by mass-binning all stars in all clusters. It is shown as
a {\it long dashed line} in Fig.~\ref{fig:mc2} for constrained
sampling and as a {\it short dashed line} for sorted
sampling. Additionally the 'input' standard stellar IMF ({\it solid
  line}) and the semi-analytical IGIMF from Fig.~\ref{fig:mc1} ({\it
  dotted line}) are shown. The IGIMF obtained with sorted sampling
agrees very well with the semi-analytical result, while the IGIMF
obtained from constrained sampling shows a flatter slope. Even
this IGIMF nevertheless differs strongly from the standard IMF with a
Salpeter slope, as a result of the imposed condition of a given
cluster mass.


\subsection{Different ECMF}
\label{subsec:ecmf}
Following-on from our discussion in \citet{WK05} we explore how a
different ECMF affects the IGIMF. Any ECMF with $\beta > 2.35$ will
increase the steepening of the IGIMF. However, below about 50 or 100
$M_{\odot}$ the ECMF is poorly defined observationally \citep{LL03},
and we consider here the implication of a flatter ECMF at low masses
while keeping $\beta = 2.35$ for $M_{\rm ecl} > 50$ or 100 $M_{\odot}$.

\begin{figure}
\begin{center}
\vspace*{-0.6cm}
\includegraphics[width=8cm]{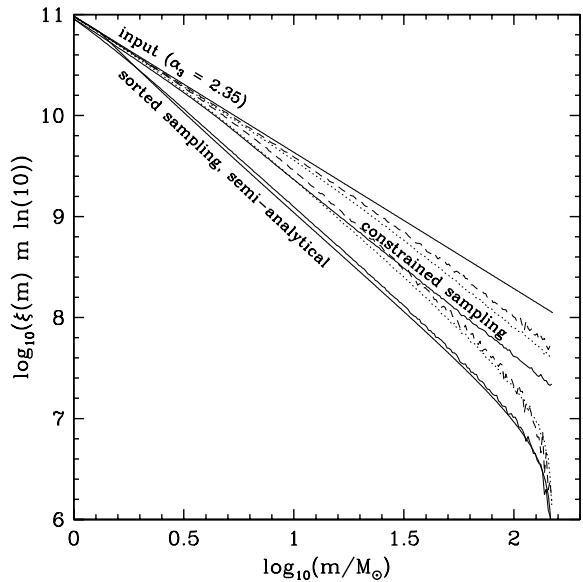}
\vspace*{-2cm}
\caption{As Fig.~\ref{fig:mc2} (all lines from Fig.~\ref{fig:mc2} 
  are {\it solid}) but in addition the results for two different ECMFs
  are plotted as {\it dotted} and {\it dashed lines}. In the {\it
    dotted} case the slope of the ECMF is $\beta_{1}\,=\,1$ for
  clusters below $50\,M_{\odot}$. In the {\it dashed} case
  $\beta_{1}\,=\,1$ below $100\,M_{\odot}$. In both cases
  $\beta_{2}\,=\,2.35$ for clusters above these limits. As the IMF
  below $1\,M_{\odot}$ does not change, only the region above
  $1\,M_{\odot}$ is shown. The respective models assuming sorted
  sampling are always steeper than those assuming constrained sampling.
\label{fig:ecmf}}
\end{center}
\end{figure}

In Fig.~\ref{fig:ecmf} we explore the effect of a different ECMF than our
standard assumption ($\beta\,=\,2.35$ over the whole range of cluster
masses). This figure is the same as Fig.~\ref{fig:mc2}, but the
constrained sampling and sorted sampling results for two EMCFs
with $\beta_{1}\,=\,1$ for clusters below $50\,M_{\odot}$ and
$100\,M_{\odot}$, and $\beta_{2}\,=\,2.35$ for clusters above these
masses, are shown as {\it dotted} and {\it dashed lines},
respectively. The steeper ones show the sorted-sampling cases. The
change of the ECMF reduces the effect of the ECMF on the IGIMF but is
still clearly visible, giving $\alpha_{\rm IGIMF, con}$ = 2.56,
$\alpha_{\rm IGIMF, sort}$ = 2.88 for $\beta = 1$ ($M_{\rm ecl} <
50\,M_{\odot}$) and $\alpha_{\rm IGIMF, con}$ = 2.52, $\alpha_{\rm
  IGIMF, sort}$ = 2.86 for $\beta = 1$ for ($M_{\rm ecl} <
100\,M_{\odot}$), as opposed to our standard assumption for which
$\alpha_{\rm IGIMF, con}\,=\,2.8$ and $\alpha_{\rm IGIMF,
  sort}\,=\,3.0$.

\section{Discussion and Conclusions}
\label{sec:concs}
In this contribution a number of Monte-Carlo experiments are conducted
in order to constrain the relation between the maximal mass a
star can have in a cluster and the mass of the cluster, and to further
study the IMF of composite populations, ie.~the integrated galaxial IMF.

We consider three possible ways of forming clusters: (1) Completely
randomly. Clusters are filled randomly with stars and then their masses,
$M_{\rm ecl}$, are calculated (random sampling). (2) Cluster-masses
are picked from an ECMF and used as a constrain in constructing the
stellar content of each cluster (constrained sampling). (3)
Cluster-masses are picked from an ECMF, and the clusters are then
filled with 
stars by randomly selecting from the canonical IMF, sorting the
stellar masses in ascending order and constraining their sum to be the
cluster mass (sorted sampling). In all cases (1-3), the most massive
star, $m_{\rm max}$, in each cluster is found, and the average or
expectation value, $\overline{m}_{\rm max}$, is calculated for the
ensemble of clusters near $M_{\rm ecl}$ to give the relations
$^{l,\,u}\overline{m}_{\rm max}^{\rm ran,\,con,\,sort}(M_{\rm ecl})$,
where ``l, u'' refers to models with or without a physical stellar
mass limit of $150\,M_{\odot}$.

The most important and surprising result is that the sorted-sampling
algorithm best represents the observational data of young ($\le$ 3
Myr) clusters. Constrained and random sampling do not fit the
observations. 

That our sorted-sampling algorithm for making stars fits the
observational maximal-stellar-mass---star-cluster-mass data so well
would appear to imply that clusters form in an {\it organised
  fashion}. The physical interpretation of the algorithm (i.e.~of the
Monte-Carlo integration) is that as a pre-cluster core contracts under
self gravity the gas densities increases and local density fluctuations
in the turbulent medium lead to low-mass star formation, perhaps
similar to what is seen in Taurus-Auriga. As the contraction proceeds
and before feedback from young stars begins to disrupt the cloud,
star-formation activity increases in further density fluctuations with
larger amplitudes thereby forming more massive stars. The process
stops when the most massive stars that have just formed supply 
sufficient feedback energy to disrupt the cloud \citep{El83}. Thus,
less-massive pre-cluster cloud-cores would ``die'' at a lower maximum
stellar mass than more massive cores. But in all cases stellar masses
are limited by the physical maximum mass, $m\,\le\,m_{\rm max}(M_{\rm
  ecl})\,\le\,m_{\rm max*}$. This scenario is nicely consistent with
the hydrodynamic cluster formation calculations presented by
\citet{BBV03} and \citet{BVB04}. 
We note here that \citet{BVB04} found their theoretical clusters to
form hierarchically from smaller sub-clusters, and together with
continued competitive accretion this leads to the relation $m_{\rm
  max}\,\propto\,M_{\rm ecl}^{2/3}$ (eq.~\ref{eq:MvMB}) in excellent
agreement with our compilation of observational data for clusters with
masses below $M_{\rm ecl}\,=\,4000\,M_{\odot}$. While this agreement
is stunning, the detailed outcome of the currently available SPH
modelling in terms of stellar multiplicities is not right
\citep{GK05}, and feedback that ultimately dominates the process of
star-formation, given the generally low star-formation efficiencies
observed in cluster-forming volumes \citep{LL03}, is not yet
incorporated in the modelling.

Stellar evolution is the major caveat here. But the comparison of
different models (see Appendix~\ref{appendixC}) shows a general
agreement of the lifetimes and relevant parameters (mass, ${\rm
  T}_{\rm eff}$ and luminosity) for the models considered
here. Therefore, not different models but an intrinsically
steeper IMF ($\alpha_{3} \simgreat 2.8$) could shift the expectation
values for random and constrained sampling into the observed
regime. Such a steep IMF may be possible if it is masked by unresolved
multiple stars, something we are investigating now.

Furthermore, the Monte-Carlo experiments ascertain the results of
\citet{KW03} and \citet{WK05} regarding the steep IGIMF, especially so if
sorted sampling is used. In the constrained sampling case the IGIMF slope
is still steeper than the input IMF but less steep than with sorted
sampling. {But it should be noted here that a very recent result by
\citet{ES05} shows that it is also possible to interpret PDMF variations
falsely, as IMF variations when the SFR is assumed to be constant when
in reality being burst-like. This result has not yet been implemented
in our description of the IGIMF.}

In summary:
\begin{itemize}
\item[-] There exists a well-defined relation, $m_{\rm max} = m_{\rm
    max}(M_{\rm ecl})$, between the most-massive star in a cluster and
  the cluster mass. The conjecture that a cluster consists of stars
  randomly picked from an invariant IMF between 0.01 and 150
  $M_{\odot}$ would therefore appear to be wrong.
\item[-] Star clusters appear to form in an ordered fashion, starting
  with the lowest-mass stars until feedback is able to outweigh the
  gravitationally-induced formation process.
\item[-] IGIMFs must {\it always} be steeper for $m>1\,M_\odot$
than the stellar IMF that results from a local star-formation
event. 
\end{itemize}

It will be important to further test the results presented here on the
$m_{\rm max}(M_{\rm ecl})$ relation by compiling larger observational
samples of young clusters. {As this contribution has shown, it appears
that the $m_{\rm max}(M_{\rm ecl})$ relation would be rather
fundamental to galactic and extragalactic astrophysics.}
\section*{Acknowledgements}

We are very thankful to Jarrod R.~Hurley for kindly providing us with his
very useful single stellar evolution (SSE) software package, and to
Simon P.~Goodwin and Jan Pflamm for useful discussions. This research
has been supported by DFG grant KR1635/3.


\bsp
\begin{appendix}
\section{cluster and maximal star mass determination}
\label{appendixA}
The masses of the clusters in Tab.~\ref{tab:clusters} are acquired as
follows:

In the case of NGC1333, NGC2024, NGC6530, M42, $\rho$ Ophiuchi,
$\sigma$ Orionis, Serpens SVS2, Arches and R136 the authors of the
corresponding papers provide the required mass estimates. The masses
for NGC2244, NGC2264, NGC6611, IC348, Monoceros R2, Taurus-Auriga, Berkley 86
and Trumpler 14/16 are calculated by {determining the fraction
  (given as a percentage in Tab.~\ref{tab:mclus})
  of observed stars in comparison with a canonical IMF from 0.01
  $M_{\odot }$ up to the observed upper mass limit. With the
  fraction of all stars the total number of stars in the cluster is
  estimated by dividing the observed number of stars by the
  fraction. The total mass of the cluster, $M_{\rm ecl}$, is then
  calculated by multiplying the total number of stars with the mean
  stellar mass, $m_{\rm mean}$, for the canonical IMF from 0.01
  $M_{\odot}$ to the observed upper mass limit. The relevant values
  are shown in Tab.~\ref{tab:mclus}.}

\begin{table}
\centering
\caption{\label{tab:mclus}
  Observed mass ranges and the corresponding mean masses {for those
  clusters for which we need to derive cluster masses}.}
\vspace*{0.5cm}
\begin{tabular}{ccccccc}
\hline
Cluster&number&\% of&total&mass range&$m_{\rm mean}$&$M_{\rm ecl}$\\
&of stars&all&number&[$M_{\odot}$]&[$M_{\odot}$]&[$M_{\odot}$]\\
&observed&stars&of stars&&&\\
\hline
Tau-Aur&100&92.8&108&0.02 - 2.2&0.236&25.5\\
IC348&241&62.5&386&0.08 - 6.0&0.282&109\\
Mon R2&475&55.2&861&0.1 - 10.0&0.301&259\\
NGC2264&600&55.3&1086&0.1 - 25.0&0.327&355\\
Ber 86&340&7.7&4421&0.8 - 40.0&0.338&1500\\
NGC2244&400&2.2&17937&2.0 - 70.0&0.348&6240\\
NGC6611&362&0.6&56563&5.0 - 85.0&0.352&20000\\
Tr 14/16&768&0.6&120000&5.0 - 120.0&0.357&43000\\
\hline
\end{tabular}
\end{table}

For the maximal stellar masses in these clusters the values within the
papers are used whenever possible, which is for NGC2244, NGC2264, NGC6611,
M42, $\rho$ Ophiuchi, $\sigma$ Orionis, Monoceros R2, Berkley 86,
Trumpler 14/16, Arches and R136. In the other cases (NGC1333, NGC2024, NGC6530,
IC348, Serpens SVS2, Taurus-Auriga) mass estimates are derived from
the spectral types of the most luminous members using the
spectral-type mass-relation from \citet{Co00}.

\section{Fitting Formulae For Massive Star Evolution}
\label{appendixB}
Because the \citet{HPT00} single stellar evolution (SSE) package is only
calibrated for stellar models up to $50\,M_{\odot}$, additional
fitting formulae have been developed for more-massive stars. Based on the
\citet{SSM92} models for 60, 85 and 120 $M_{\odot}$, functions for
$m(t)$, $L(t)$ and $T_{\rm eff}(t)$ have been obtained:

\subsection{Mass Evolution}
As long as the age, $t$ (in Myr), of the star is below $\tau_{\rm m}$,
the main-sequence life-time, the mass-evolution can be described
according to,
\begin{equation}
m(t) = m_{\rm ini} \cdot e^{-(a_{1}\cdot t)^{2}}.
\end{equation}

For $\tau_{\rm m}\,<\,t\,\le\,\tau_{\rm m}\,+\,dt$ 
\begin{equation}
m(t) = -\frac{a_{2}}{dt} \cdot t + b_{1}.
\end{equation}

For both the parameters are
\begin{eqnarray*}
\tau_{\rm m} &=& \frac{\left(e^{-\frac{m_{\rm ini}}{t_{af}}\cdot
      t_{am}}\right)+t_{ab}}{10^{6}},\\
a_{1} &=& \frac{1}{\left(m_{aa} \cdot e^{-\frac{m_{\rm
          ini}}{m_{ab}}}\right)+m_{ac}},\\
a_{2} &=& m_{\rm ini} \cdot (1-f_{1}),\\
f_{1} &=& (f_{m} \cdot m_{\rm ini}) - f_{b},\\
b_{1} & = & a_{2} \cdot \left(1 + \frac{1}{dt} - \frac{\tau_{\rm m}}{a_{2}}\right),
\end{eqnarray*}
and constants
\begin{eqnarray*}
m_{aa}& = &48.0,\\
m_{ab}& = &24.7,\\
m_{ac}& = &3.15,\\
dt& = &0.42,\\
f_{m}& = &3.523808 \cdot\,10^{-3},\\
f_{b}& = &6.190428 \cdot\,10^{-3},\\
t_{af}& = &23.5,\\
t_{am}& = &1.25 \cdot\,10^{7},\\
t_{ab}& = &2.5 \cdot\,10^{6}.
\end{eqnarray*}
When $t$ is larger than $\tau_{\rm m} + dt$ the star is considered
dead. No remnant mass, $T_{\rm eff}$ or luminosity is assigned.

\begin{figure}
\begin{center}
\vspace*{-0.6cm}
\includegraphics[width=8cm]{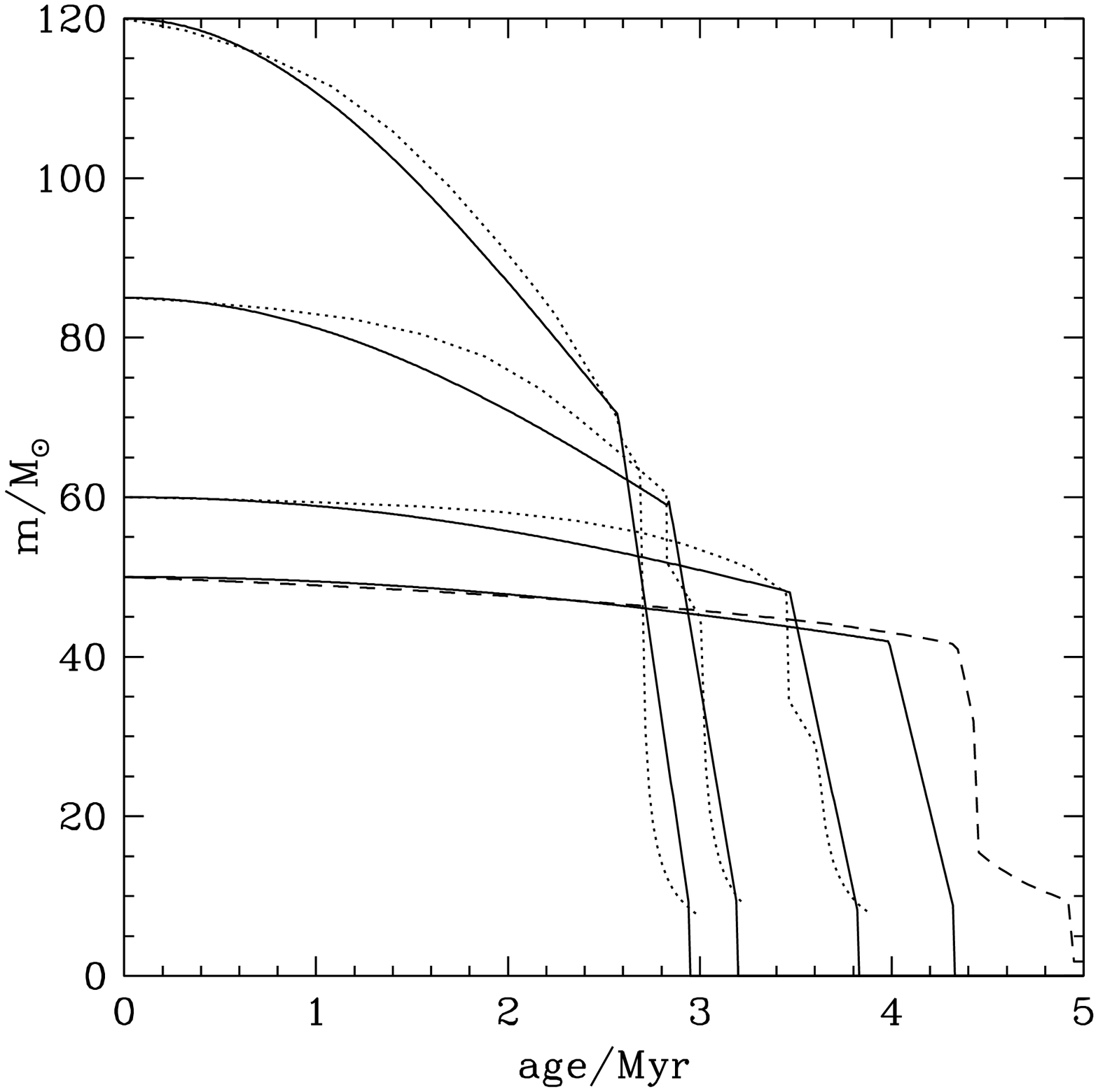}
\vspace*{-2cm}
\caption{Stellar mass
  evolution for massive stars. The {\it dotted lines} are the Geneva
  models \citep{SSM92}, while the {\it solid lines} are the results
  from the fitting formulae described here. A \citet{HPT00}
  $50\,M_{\odot}$-star is shown as a {\it dashed line}.
\label{fig:Fitmass}}
\end{center}
\end{figure}

The resulting curves for a 120, 85, 60 and
50 $M_{\odot}$ star ({\it solid lines}) in comparison with the model
data ({\it dotted lines}) are plotted in Fig.~\ref{fig:Fitmass}.

\subsection{Effective Temperature Evolution}
For $t\,\le\,\tau_{\rm m}$ the equation
\begin{equation}
T_{\rm eff}(t) = T_{\rm eff,\, ini} \cdot e^{\left(-\frac{t}{a_{3}}\right)^{b_{2}}},
\end{equation}
adequately captures the evolution.

\noindent
For $\tau_{\rm m}\,<\,t\,\le\,\tau_{\rm m}\,+\,dt$,
\begin{equation}
T_{\rm eff}(t) = f_{2} \cdot t + (T_{\rm low} -
(f_{2}\,\cdot\,\tau_{\rm m})),
\end{equation}
is used.\\

The parameters are
\begin{eqnarray*}
T_{\rm eff,\,ini} &=& \left\{ \begin{array}{ll}
T_{\rm im_{1}} \cdot m_{\rm ini} + T_{\rm ib_{1}}&m_{\rm ini} \ge 60
M_{\odot},\\
T_{\rm im_{2}} \cdot m_{\rm ini} + T_{\rm ib_{2}}&m_{\rm ini} < 60 M_{\odot},
\end{array} \right.\\
a_{3} &=& \left\{ \begin{array}{ll}
tta_{1}&m_{\rm ini} \ge 85 M_{\odot},\\
tta_{2\rm a} \cdot m_{\rm ini} + tta_{2\rm b}&m_{\rm ini} < 85 M_{\odot},
\end{array} \right.\\
f_{2}&=& \frac{(T_{\rm peak} - T_{\rm low})}{dt},\\
T_{\rm low} &=& T_{\rm eff,\,ini} \cdot e^{\left(-\frac{\tau_{\rm
        m}}{a_{3}}\right)^{b_{2}}},
\end{eqnarray*}
and the constants
\begin{eqnarray*}
T_{\rm im_{1}}&=&4.3\,\cdot\,10^{-3},\\
T_{\rm ib_{1}}&=&4.425,\\
T_{\rm im_{2}}&=&7.3333\,\cdot\,10^{-4},\\
T_{\rm ib_{2}}&=&4.64,\\
b_{2}&=&2.3,\\
tta_{1}&=&9.0,\\
tta_{2\rm a}&=&-0.182857142,\\
tta_{2\rm b}&=&24.54285719,\\
T_{\rm peak}&=&4.8.
\end{eqnarray*}

\begin{figure}
\begin{center}
\vspace*{-0.6cm}
\includegraphics[width=8cm]{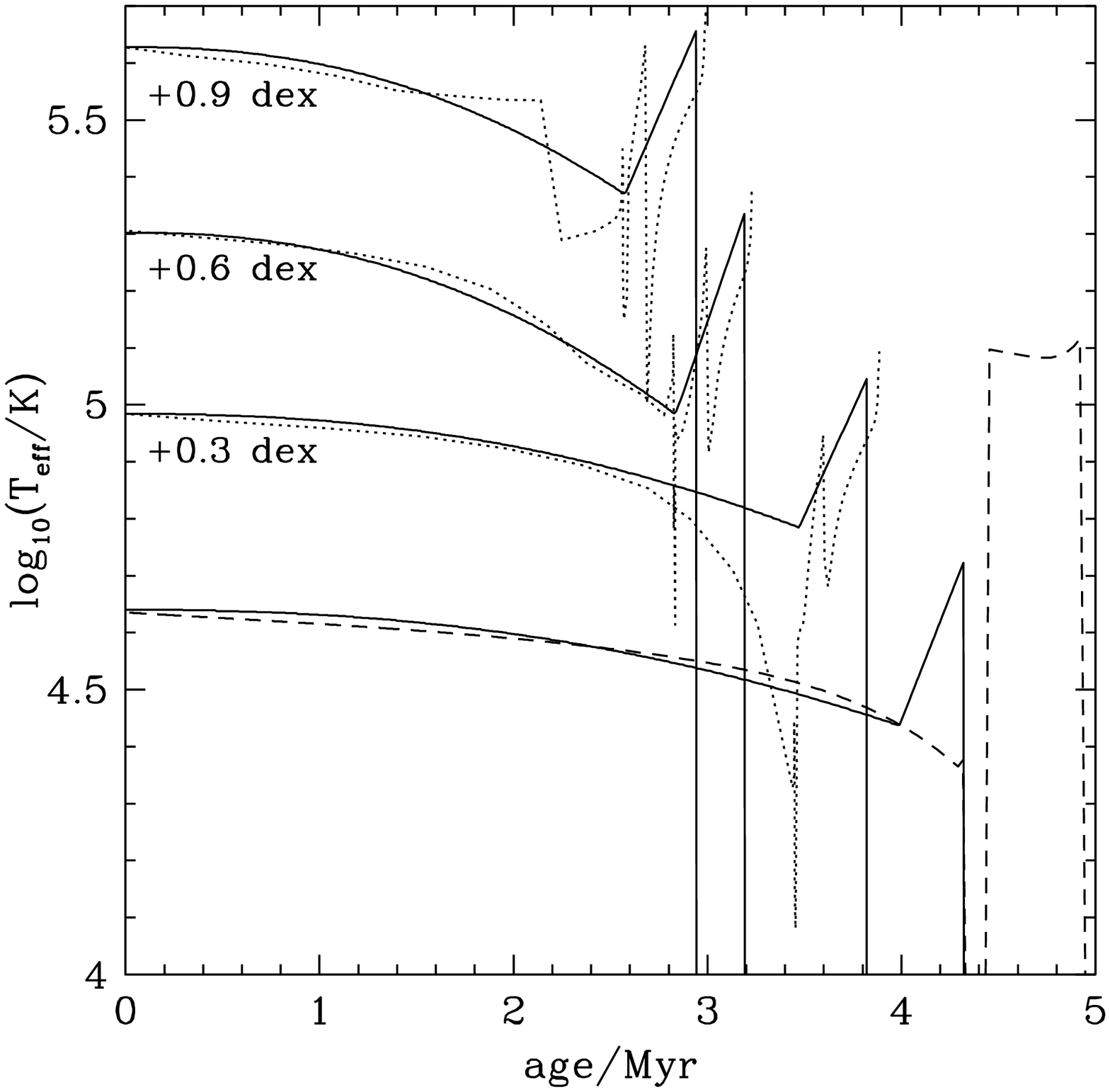}
\vspace*{-2cm}
\caption{Effective
  temperature evolution for massive stars. The line styles are as in
  Fig.~\ref{fig:Fitmass}. As the $T_{\rm eff}$ for massive stars are
  rather similar, the lines in the plot have been shifted upwards as
  indicated in the plot.
\label{fig:Fitteff}}
\end{center}
\end{figure}

The resulting $T_{\rm eff}$ fitting curves are plotted in
Fig.~\ref{fig:Fitteff} for the same masses as in Fig.~\ref{fig:Fitmass}.

\subsection{Luminosity Evolution}
The luminosity evolution is divided into three parts. For
$t\,\le\,\tau_{\rm m}$,
\begin{equation}
L(t) = \left(\frac{L_{\rm jump} - L_{\rm ini}}{\tau_{\rm m}}\right) \cdot t +
L_{\rm ini}.
\end{equation}

\noindent
For $\tau_{\rm m} < t < \tau_{\rm b}$
\begin{equation}
L(t) = L_{\rm jump} + (L_{\rm peak} - L_{\rm jump}) \cdot \sin(L_{\rm
sin} \cdot (t - \tau_{\rm m})).
\end{equation}

\noindent
And finally for $\tau_{\rm b}\,\le\,t\,\le\,(\tau_{\rm m} + dt)$
\begin{equation}
L(t) = a_{4} \cdot t + b_{3}.
\end{equation}

\noindent
Here the parameters are
\begin{eqnarray*}
L_{\rm ini}&=&\frac{m_{\rm ini}}{M_{L}},\\
M_{L}&=&ml_{\rm m} \cdot m_{\rm ini} + ml_{\rm b},\\
L_{\rm jump}&=&\exp\left[{\frac{L_{\rm je}}{\left(L_{\rm ini}^{L_{\rm
          pf}}\right)}}\right] + L_{\rm jc},\\
\tau_{\rm b}&=&\frac{1\,\cdot\,10^{-6}}{\tau_{\rm bb} -
  (\frac{\tau_{\rm bm}}{m_{\rm ini}})},\\
L_{\rm peak}&=&L_{\rm pa} \cdot \log_{10}\left(\frac{m_{\rm ini}}{L_{\rm
    pf}}\right) + L_{\rm pc},\\
a_{4}&=&\frac{L_{\rm jump} - L_{\rm low}}{\tau_{\rm m} + dt -
  \tau_{b}},\\
b_{3}&=&L_{\rm low} - (a_{4} \cdot (\tau_{\rm m} + dt)).
\end{eqnarray*}
The constants are
\begin{eqnarray*}
ml_{\rm m}&=&0.145590532,\\
ml_{\rm b}&=&1.722994092,\\
L_{\rm je}&=&9.0\,\cdot\,10^{-3},\\
L_{\rm pf}&=&2.5,\\
L_{\rm jc}&=&3.95,\\
\tau_{\rm bb}&=&4.6296293\,\cdot\,10^{-7},\\
\tau_{\rm bm}&=&1.1111\,\cdot\,10^{-5},\\
L_{\rm sin}&=&15.7,\\
L_{\rm pa}&=&1.35,\\
L_{\rm pc}&=&4.215,\\
L_{\rm low}&=&5.2.
\end{eqnarray*}

\begin{figure}
\begin{center}
\vspace*{-0.6cm}
\includegraphics[width=8cm]{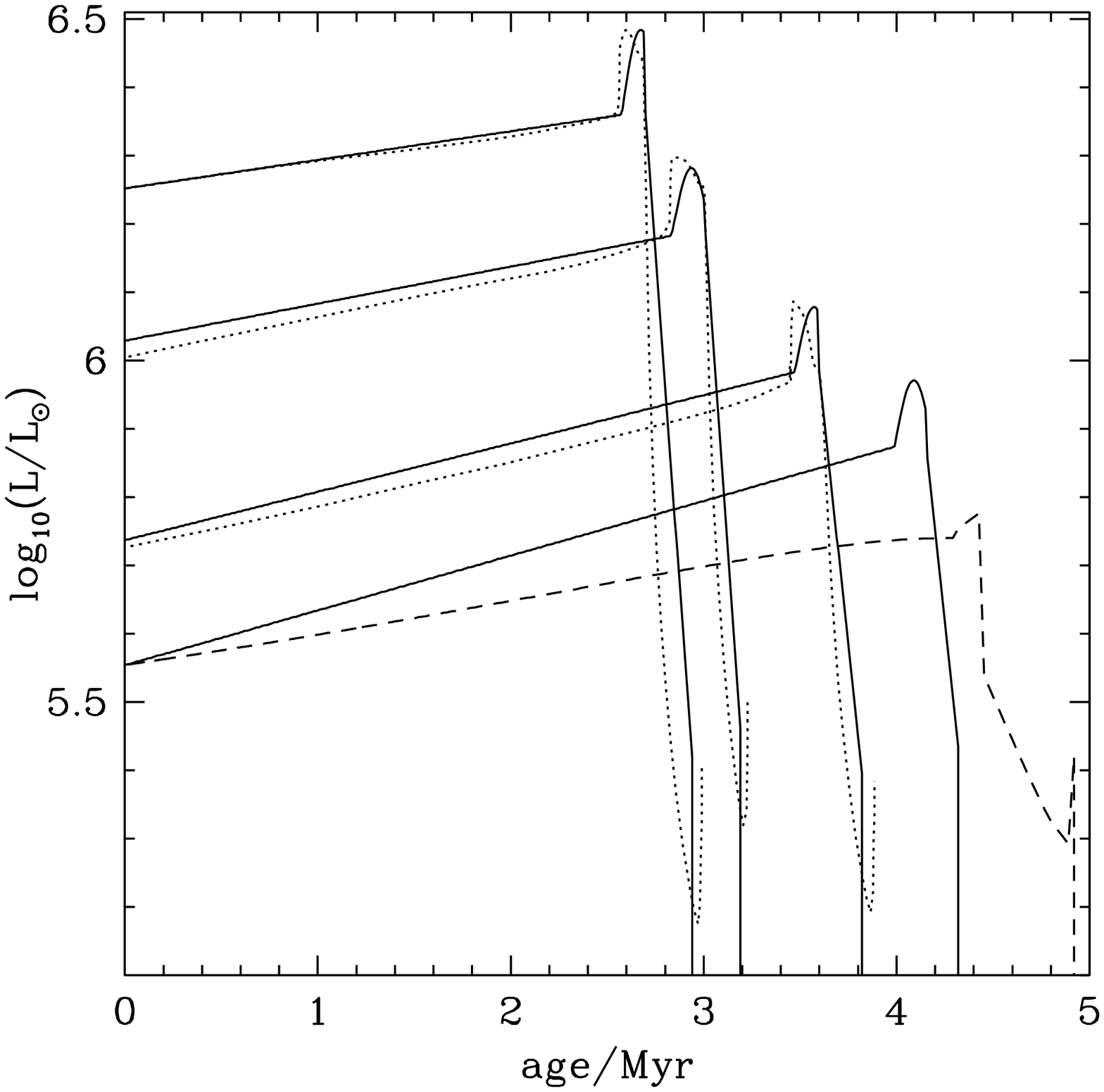}
\vspace*{-2cm}
\caption{Luminosity evolution
  for massive stars. The line styles are as in Fig.~\ref{fig:Fitmass}.
\label{fig:Fitlum}}
\end{center}
\end{figure}

Fig.~\ref{fig:Fitlum} shows the results of the luminosity-fitting formulae.

\section{Comparison of different models}
\label{appendixC}
\begin{figure}
\begin{center}
\vspace*{-0.6cm}
\includegraphics[width=8cm]{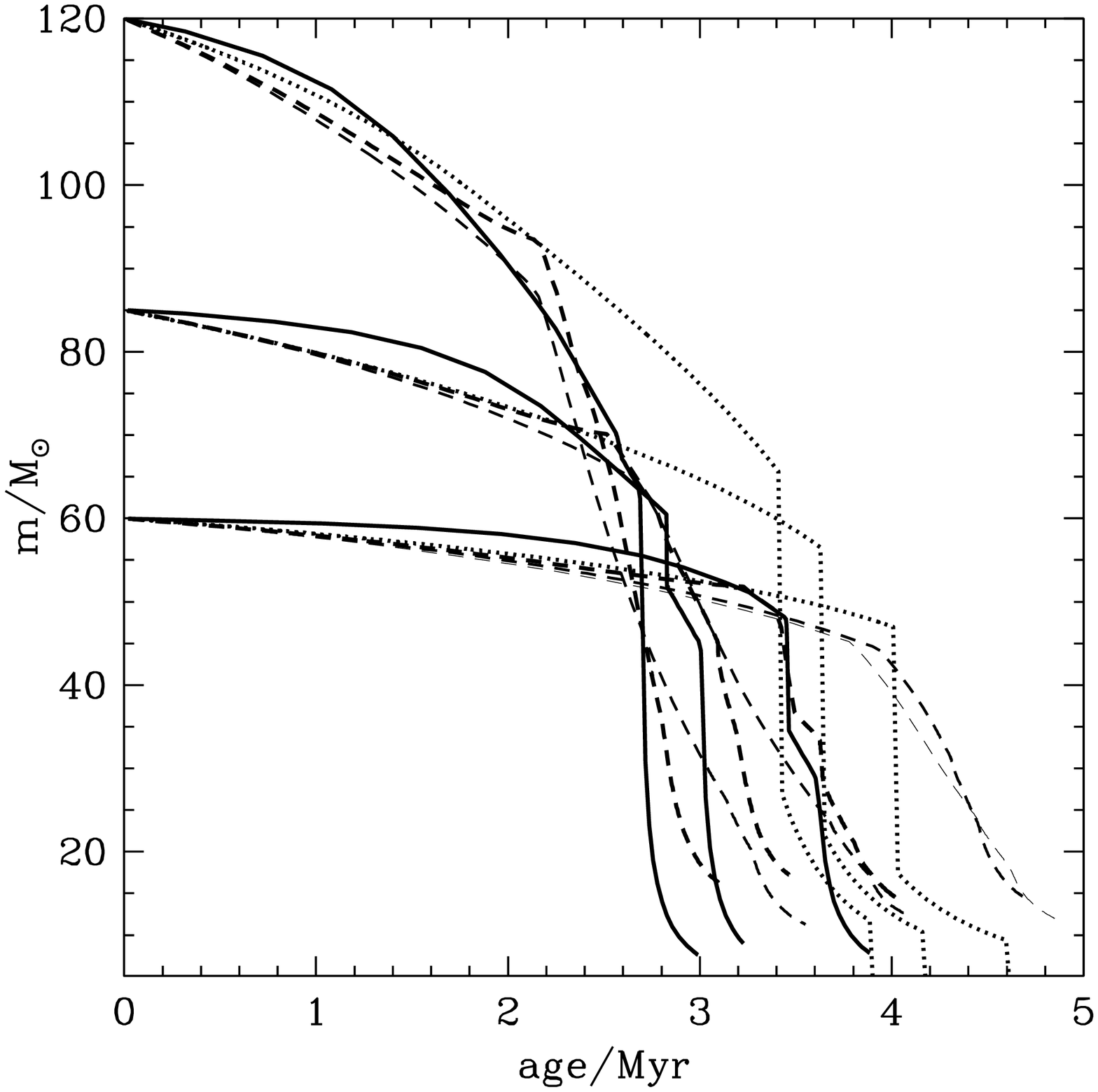}
\vspace*{-2cm}
\caption{Stellar mass evolution for massive stars. The {\it solid
    lines} are the Geneva models \citep{SSM92}, while the {\it dashed
    lines} are the Geneva models with rotation \citep[][{\it thick
    dashed line}: no rotation, {\it medium dashed line}: 300 km/s,
  {\it thin dashed line}: 500 km/s, only for
  $m\,=\,60\,M_{\odot}$]{MM03} and the results from the SSE
  package \citep{HPT00} are shown as {\it dotted lines}.
\label{fig:compmass}}
\end{center}
\end{figure}

\begin{figure}
\begin{center}
\vspace*{-0.6cm}
\includegraphics[width=8cm]{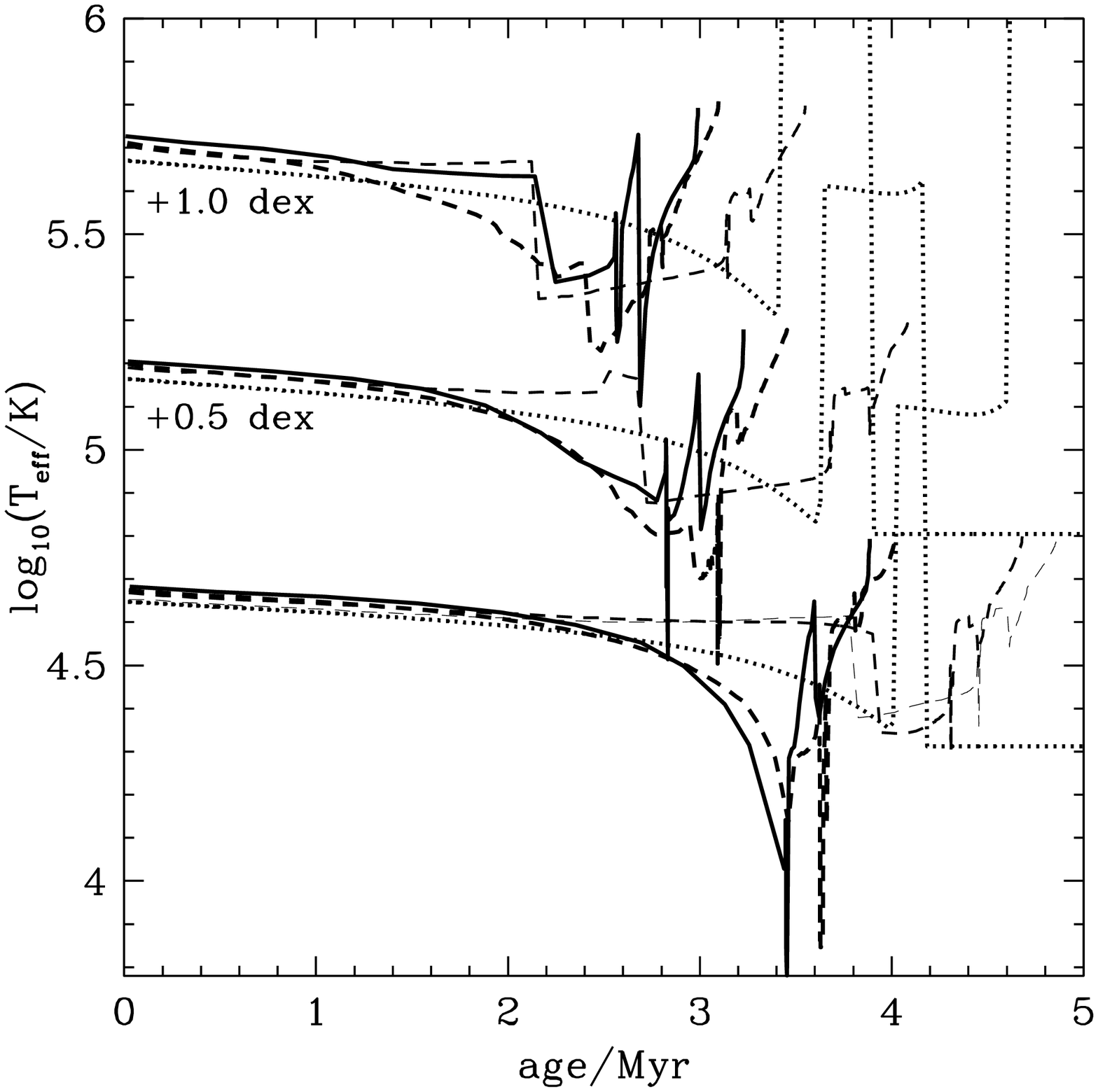}
\vspace*{-2cm}
\caption{Effective temperature evolution for massive stars. The line
  styles are as in Fig.~\ref{fig:compmass}. As the $T_{\rm eff}$ for
  massive stars are rather similar, the lines in the plot have been
  shifted (as indicated in the plot) in the following way: the
  upper-most by +1.0 dex and the second one by +0.5 dex. The lowest
  plot {has not been} shifted.
\label{fig:compteff}}
\end{center}
\end{figure}

\begin{figure}
\begin{center}
\vspace*{-0.6cm}
\includegraphics[width=8cm]{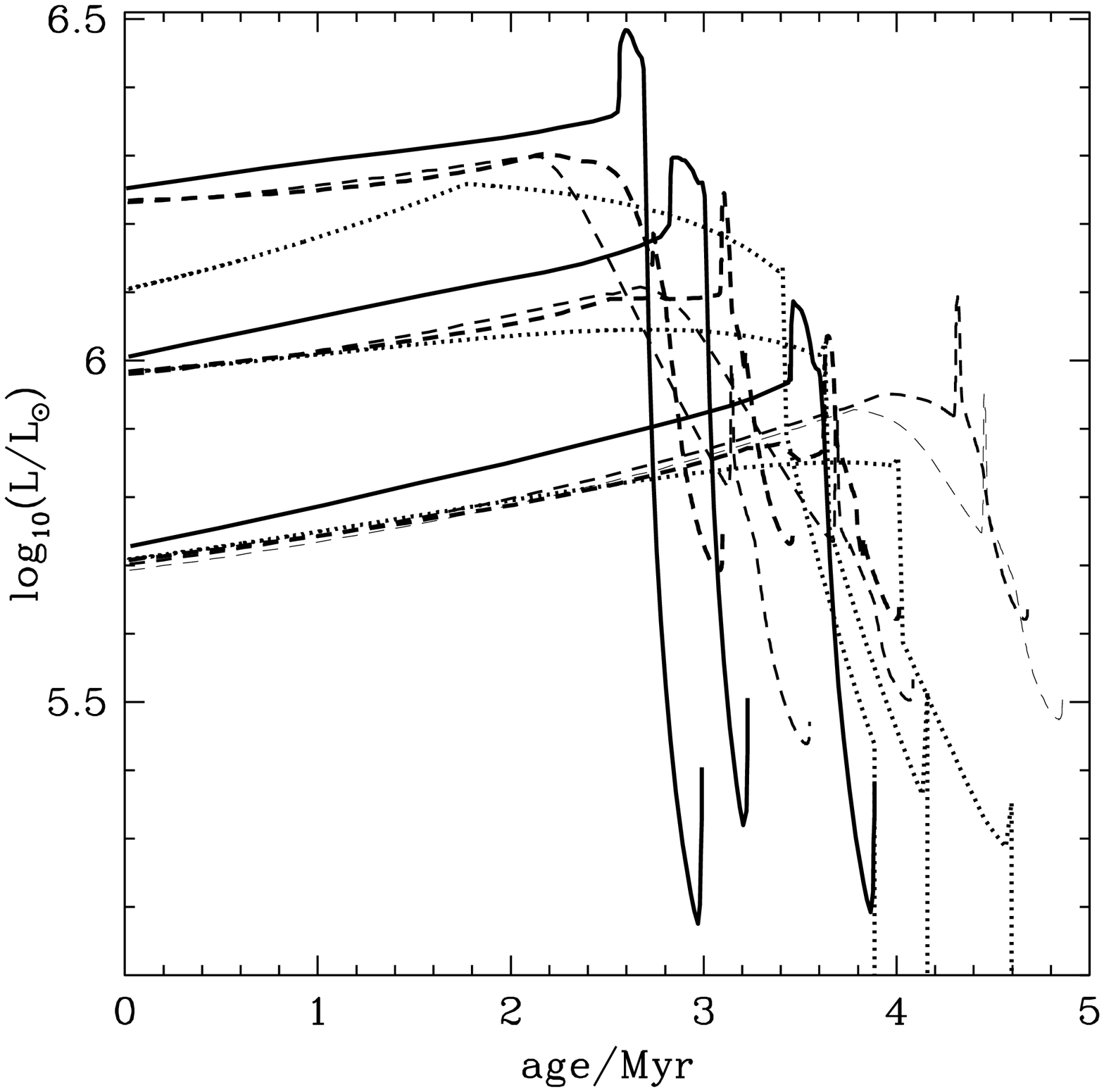}
\vspace*{-2cm}
\caption{Luminosity evolution for massive stars. The line styles are as
  in Fig.~\ref{fig:compmass}.
\label{fig:complum}}
\end{center}
\end{figure}

Several different sets of theoretical models for stellar evolution of
massive stars exist. Figs.~\ref{fig:compmass}, \ref{fig:compteff}
and \ref{fig:complum} compare the mass, $T_{\rm eff}$ and luminosity
evolution of three different sets of models
\citep{SSM92,MM03,HPT00}. These models agree qualitatively on the
compared stellar properties but show minor differences in the
details.
\end{appendix}
\label{lastpage}
\end{document}